\documentclass[reprint,amsmath,amssymb,aps,nofootinbib,superscriptaddress]{revtex4-2}

\usepackage[utf8]{inputenc}
\usepackage{graphicx}
\usepackage{soul}
\usepackage{xcolor}
\usepackage{amsmath}
\usepackage{amsfonts}
\usepackage{latexsym}
\usepackage{amssymb,bbm}
\usepackage{bm}
\usepackage[colorlinks=true,allcolors=blue]{hyperref}
\usepackage{setspace}
\usepackage{braket}

\newcommand{\beq}{\begin{equation}}
\newcommand{\eeq}{\end{equation}}
                  
\newcommand{\benum}{\begin{enumerate}}
\newcommand{\eenum}{\end{enumerate}}
                    
\newcommand{\bit}{\begin{itemize}}
\newcommand{\eit}{\end{itemize}}

\newcommand{\bea}{\begin{eqnarray}}
\newcommand{\eea}{\end{eqnarray}}

\newcommand{\bev}{\begin{verbatim}}
\newcommand{\eev}{\end{verbatim}}

\newcommand{\zfl}[1]{\protect\label{fig:#1}}

\newcommand{\ba}{\left\{ \begin{array}{lr}}
\newcommand{\ea}{\end{array}\right.}

\newcommand{\blist}[1]{
 \begin{list}{#1}
 \begin{align}
	 arrow
 \end{align}
 $\checkmark\star
  { \setlength{\itemsep}{3pt}
     \setlength{\parsep}{2pt}
     \setlength{\topsep}{3pt}
     \setlength{\partopsep}{0pt}
     \setlength{\leftmargin}{1em}
     \setlength{\labelwidth}{1em}
     \setlength{\labelsep}{0.5em} } }
\newcommand{\elist}{
  \end{list}  }

\DeclareMathSymbol{\vartheta}{\mathalpha}{letters}{"12}
\DeclareMathSymbol{\theta}{\mathalpha}{letters}{"23}
\DeclareMathSymbol{\phi}{\mathalpha}{letters}{"27}
\DeclareMathSymbol{\varphi}{\mathalpha}{letters}{"1E}

\newcommand{\bef}
{
\begin{figure}[htbp]
\centering
}

\newcommand{\eef}{\end{figure}}

\makeatletter
\newcommand{\beginsupplement}{%
  \clearpage
  \setcounter{section}{0}%
  \setcounter{subsection}{0}%
  \setcounter{equation}{0}%
  \setcounter{figure}{0}%
  \setcounter{table}{0}%
  \renewcommand{\thesection}{S\arabic{section}}%
  \renewcommand{\thesubsection}{\Alph{subsection}}%
  \renewcommand{\p@subsection}{\thesection~}%
  \renewcommand{\p@subsubsection}{}%
  \renewcommand{\theequation}{S\arabic{equation}}%
  \renewcommand{\thefigure}{S\arabic{figure}}%
  \renewcommand{\thetable}{S\arabic{table}}%
}
\makeatother

\begin{document}

\title{Breakdown of Disorder-Suppressed Floquet Heating under Two-Frequency Driving}

\author{Cooper M. Selco}
\thanks{These authors contributed equally to this work.}
\affiliation{Department of Chemistry, University of California, Berkeley, Berkeley, CA 94720, USA}

\author{Christian Bengs}
\thanks{These authors contributed equally to this work.}
\affiliation{Department of Chemistry, University of California, Berkeley, Berkeley, CA 94720, USA}
\affiliation{Chemical Sciences Division, Lawrence Berkeley National Laboratory, Berkeley, CA 94720, USA}
\affiliation{School of Chemistry, University of Southampton, Southampton, SO17 1BJ, UK}

\author{Chaitali Shah}
\affiliation{Department of Chemistry, University of California, Berkeley, Berkeley, CA 94720, USA}

\author{Ashok Ajoy}
\email{ashokaj@berkeley.edu}
\affiliation{Department of Chemistry, University of California, Berkeley, Berkeley, CA 94720, USA}
\affiliation{Chemical Sciences Division, Lawrence Berkeley National Laboratory, Berkeley, CA 94720, USA}

\begin{abstract}
Periodic (Floquet) driving enables Hamiltonian engineering and nonequilibrium phases, but interacting systems eventually heat by absorbing energy from the drive. Disorder can greatly delay this process, yielding long-lived prethermal plateaus. Here we show that this protection can fail when pulse-train control introduces a second driving frequency and when the disorder fluctuates. Using a natural-abundance $^{13}$C nuclear-spin network in diamond, we observe sharp peaks in the late-time heating rate at the double- and triple-spin-flip resonance conditions predicted by bimodal Floquet interference, and track their evolution with drive frequency. A switching-noise model attributes the resonant absorption to stochastic electron-spin dynamics that intermittently tune rare nuclear clusters into multi-photon resonance. Our results reveal a resonance-activated limit for disorder-stabilized Floquet phases and suggest new routes to DC-field quantum sensing based on an abrupt breakdown of prethermalization.

\end{abstract}

\maketitle

\begin{figure*}[t]
    \centering
    \includegraphics[width=1\textwidth]{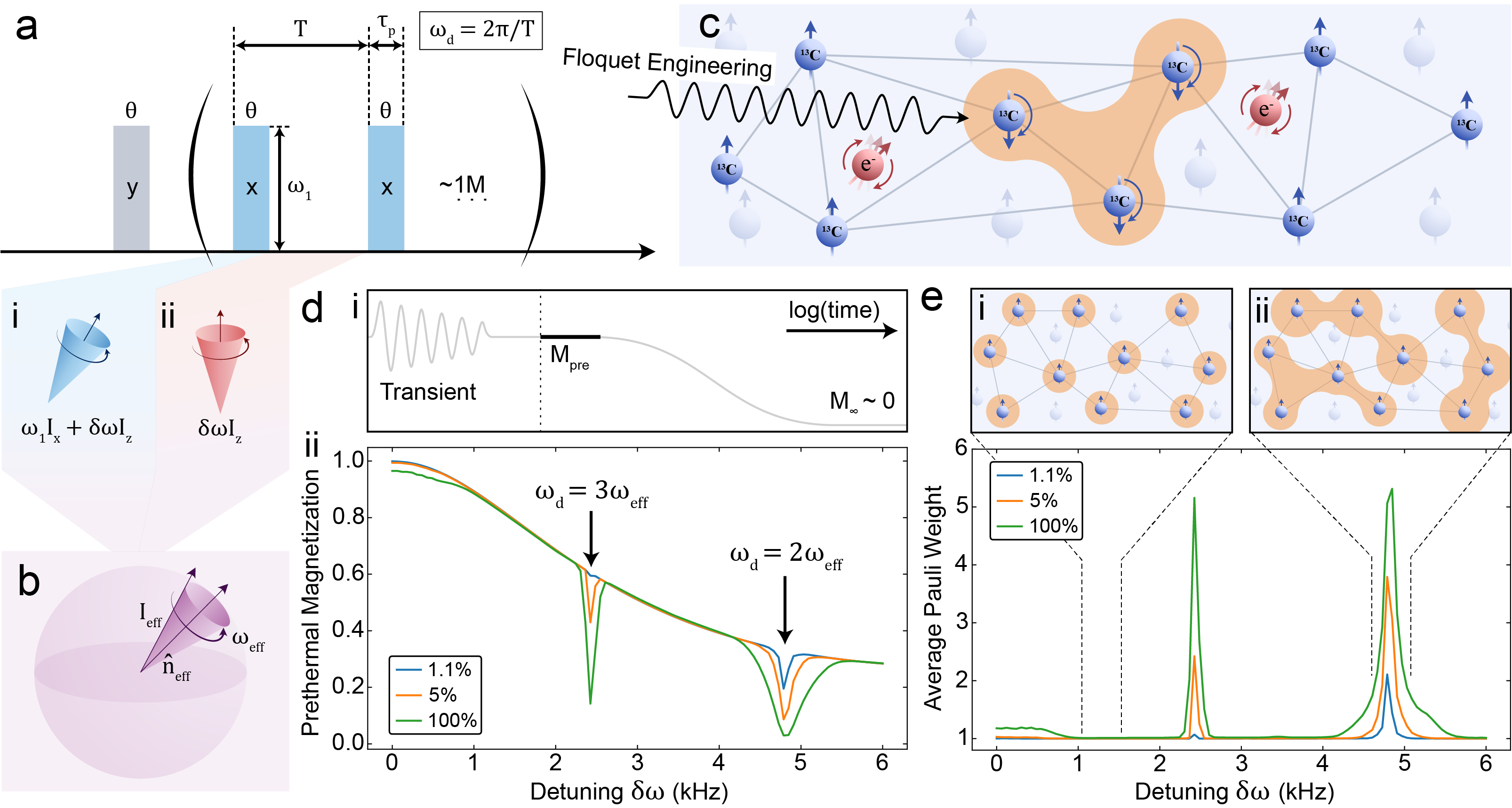}
    \caption{(a) Floquet driving sequence: initial $\vartheta_y$ pulse followed by train of detuned $\vartheta_x$ pulses (flip angle $\vartheta=\omega_{1}\tau_{p}$, detuning $\delta\omega$, inter-pulse spacing $\tau_s$) with period $T$. During pulse (i) $H_{\rm drive}=\omega_{1} I_x+\delta\omega I_z$; during delay (ii) $H_{\rm drive}=\delta\omega I_z$. (b) Net rotation over one period defining $\hat{n}_{\rm eff}$ and $\omega_{\rm eff}$. (c) Schematic of positionally disordered $^{13}$C dipolar network coupled to randomly distributed electron spins (red), illustrating triple-resonance condition ($3\omega_{\rm eff}=\omega_d$) leading to triple-spin-flip. (d) Prethermal magnetization vs. $\delta\omega$: (i) definition of $M_{\rm pre}=\sqrt{\langle I_x\rangle^{2}+\langle I_y\rangle^{2}}$, where $\langle I_\mu\rangle={\rm Tr}[I_\mu\rho_{\rm pre}]$; (ii) numerics for random 10-spin cluster on diamond lattice at 1.1\%, 5\%, and 100\% $^{13}$C occupancy, showing resonances near 2.5 kHz ($k=3$) and 4.9 kHz ($k=2$). (e) Average Pauli weight of prethermal state, with resonance-enhanced operator spreading; insets show schematic operator support far from (i) and near (ii) resonance. Simulations performed with $\tau_{p}=56$ $\mu$s, $T=92$ $\mu$s and $\omega_{1}=4.46$ kHz.}
    \label{Fig1}
    \zfl{fig1}
\end{figure*}

\begin{figure}[t]
    \centering
    \includegraphics[width=\columnwidth]{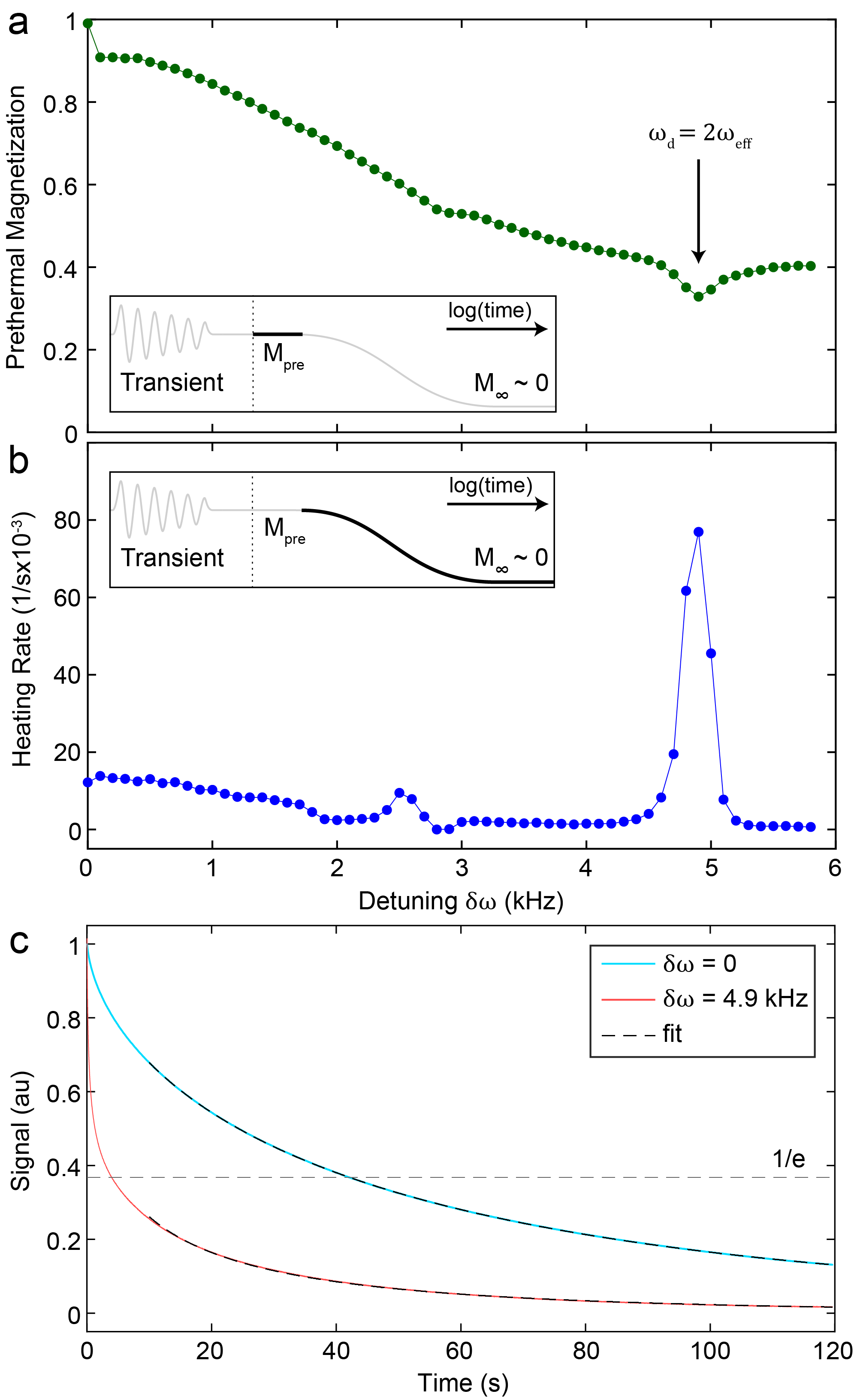}
    \caption{(a) Experimental prethermal magnetization $M_{\rm pre}$ vs. detuning $\delta\omega$ obtained by integrating signal from 10--20 ms. Feature near 4.9 kHz corresponds to double-spin-flip resonance; transient effects are small due to weak dipolar network. (b) Long-time heating rate vs. $\delta\omega$ (obtained from fits to decay, see SI Sec.~\ref{sec:SI_fitting}), showing pronounced peaks at triple- and double-spin-flip resonances. (c) Representative decays of $M_{\rm pre}(t)$ away from resonance ($\delta\omega=0$) and near double-spin-flip resonance ($\delta\omega=4.9$ kHz); dashed lines are fits used to extract heating rates (see SI Sec.~\ref{sec:SI_fitting}). Data taken with $\tau_{p}=56$ $\mu$s, $T=92$ $\mu$s and $\omega_{1}=4.46$ kHz.}
    \label{Fig2}
    \zfl{fig2}
\end{figure}



\emph{Introduction---}Periodic (Floquet) driving enables the synthesis of effective Hamiltonians and dynamical phases with no equilibrium analogue~\cite{dalessioLongtimeBehaviorIsolated2014, moriThermalizationPrethermalizationIsolated2018, okaFloquetEngineeringQuantum2019a}. Such driven Hamiltonian engineering underpins quantum simulation and coherent control across diverse platforms~\cite{lindnerFloquetTopologicalInsulator2011, rechtsmanPhotonicFloquetTopological2013, wangObservationFloquetBlochStates2013, rudnerBandStructureEngineering2020, weitenbergTailoringQuantumGases2021, randallManybodylocalizedDiscreteTime2021a, parkSteadyFloquetAndreev2022, zhouPseudospinselectiveFloquetBand2023, peng2021floquet, martin2023controlling}. A generic limitation is heating: in interacting systems, absorption from the drive ultimately erases structure and drives the state toward an effectively infinite-temperature ensemble~\cite{kuwaharaFloquetMagnusTheory2016, moessner2017equilibration, abaninEffectiveHamiltoniansPrethermalization2017, hoQuantumClassicalFloquet2023}. In practice, long-lived prethermal regimes can emerge when heating is parametrically slow, for example when the drive is fast~\cite{scholzOperatorbasedFloquetTheory2010a, bukovUniversalHighfrequencyBehavior2015, eckardtHighfrequencyApproximationPeriodically2015b, abaninExponentiallySlowHeating2015, ivanovFloquetTheoryMagnetic2021} and/or when disorder inhibits resonant absorption and transport~\cite{serbynLocalConservationLaws2013a, yaoManyBodyLocalizationDipolar2014a, smithManybodyLocalizationQuantum2016a}. A central question for Floquet engineering is therefore not only whether a prethermal plateau forms, but what ultimately sets its lifetime in realistic experiments~\cite{hou2025floquet, guo2025dynamical}.
\newline\indent
Most theoretical treatments of disorder-suppressed Floquet heating make two simplifying assumptions: (i) the drive is effectively single-frequency, and (ii) the disorder is static~\cite{ponteManyBodyLocalizationPeriodically2015, ZhangFloquetModelLocalizationTransition2016}. Both assumptions are routinely violated. In many solid-state platforms, on-site disorder is not purely static but originates from a surrounding dynamical environment. A static mean-field description is insufficient when these degrees of freedom are spatially disordered and intrinsically time-dependent, leading to stochastic site-dependent fields~\cite{joos2022protecting, davisProbingManybodyDynamics2023a, psaroudakiSkyrmionQubitsChallenges2023, selco2025emergent}. Additionally, common pulse-train and digitally stroboscopic protocols generate an additional frequency scale set by the net single-spin rotation per period, $\omega_{\mathrm{eff}}$. The resulting bimodal Floquet structure can admit discrete multi-photon resonance conditions; however, in a dilute, positionally disordered dipolar network, such higher-order processes are expected to be strongly suppressed. This raises the question of how robust disorder-based protection remains once multi-frequency driving and fluctuating disorder are both present.
\newline\indent
Here we address this question by combining bimodal Floquet theory, numerical simulations, and experiments on a natural-abundance $^{13}$C dipolar network in diamond coupled to a bath of NV and P1 electron spins. We identify double- and triple-spin-flip resonances and show that while they are less visible in the transient prethermal response, they produce pronounced, sharply peaked enhancements of the \emph{late-time} heating rate when the resonance condition is met. We experimentally map the resonance locations and strengths 
and find good agreement with the bimodal resonance conditions. We trace the breakdown of disorder protection to stochastic electron-spin dynamics: random switching of hyperfine fields intermittently tunes rare local nuclear clusters into resonance, activating otherwise suppressed many-body absorption pathways.
\newline\indent\emph{Experimental platform---}Experiments are performed on a single-crystal diamond with natural-abundance $^{13}$C (1.1\%), containing NV and P1 centers at concentrations of $\sim$1 ppm and $\sim$30 ppm, respectively~\cite{selco2025emergent}. Measurements are carried out at room temperature in a magnetic field of \mbox{$B_0=7.3$ T}. The $^{13}$C spins interact via their dipolar coupling: $H_{\rm dd}=\sum_{j<k} d_{jk}\!\left(3I_{j}^{z}I_{k}^{z}-\vec{I}_{j}\!\cdot\!\vec{I}_{k}\right)$ with $d_{jk}\propto \gamma_{\rm n}^{2}\!\left(3\cos^{2}(\vartheta_{jk})-1\right) r_{jk}^{-3}$, where $\gamma_{\rm n}=10.7\,{\rm MHz/T}$ is the gyromagnetic ratio, $r_{jk}$ is the interspin distance, and $\vartheta_{jk}$ is the angle between the internuclear vector and the magnetic field~\cite{abragam1961principles}. The crystal orientation ($\vec{B}_0\!\parallel\![100]$) is chosen to eliminate one-bond $^{13}$C couplings~\cite{selco2025emergent}; together with the low gyromagnetic ratio and $^{13}$C positional disorder, this yields a weakly coupled dipolar network with an average nearest-neighbor coupling of $\sim$100 Hz (Fig.~\ref{Fig1}c)~\cite{selco2025emergent, ajoy2019hyperpolarized}. On-site disorder originates from random hyperfine fields produced by NV and P1 centers, $H_{\rm hf}=\sum_{j,\mu} h_{j\mu} I_{j}^{z} S_{\mu}^{z}$~\cite{dreau2012high}. Measurements are initialized by $^{13}$C hyperpolarization at low magnetic field (36 mT) via polarization transfer from nearby NV centers~\cite{ajoyOrientationindependentRoomTemperature2018b, pillai2023electron}, followed by mechanical shuttling of the sample to high field where Floquet driving is performed. The hyperpolarization stage ($\sim 60$ s) is sufficiently long to produce an approximately uniform polarization via spin diffusion. In addition, the shuttling stage ($\sim 1$ s) is short compared to the nuclear $T_1$. We therefore do not expect the initialization process to introduce any effects that could appreciably influence the observed heating dynamics. Spins are driven at frequency $\omega_d=2\pi/T$ by a train of equally spaced $x$ pulses with variable detuning $\delta\omega$; the nominal flip angle is $\vartheta=\omega_{1}\tau_{p}$, where $\omega_{1}$ is the field amplitude and $\tau_{p}$ is the pulse duration (Fig.~\ref{Fig1}a)~\cite{beatrezFloquetPrethermalizationLifetime2021b}. After a transient period ($\sim T_{2}^{*}\approx1.5$ ms), a prethermal state $\rho_{\rm pre}\!\sim\! \exp(-\beta\overline{H}^{F})$ is established, where $\overline{H}^{F}$ is the Floquet Hamiltonian~\cite{beatrezFloquetPrethermalizationLifetime2021b}. Heating dynamics are monitored quasi-continuously between pulses via inductive detection of the prethermal transverse magnetization ($M_{\rm pre}$)~\cite{sahin2025micromotion}, which in the long-time limit decays to its infinite temperature value $M_{\infty}\sim0$. 
\newline\indent
\emph{Two-frequency Floquet framework---}In contrast to previous work, we focus on the off-resonant driving regime, where pulses are deliberately detuned by $\delta\omega$ in the drive Hamiltonian $H_{\rm drive}(t)\!=\!\delta\omega I_{z}+\omega_{1}(t)I_{x}$. Importantly, despite detuning, the drive remains periodic: $H_{\rm drive}(t+T)=H_{\rm drive}(t)$. During each pulse (Fig.~\ref{Fig1}a(i)), spins rotate around a tilted axis set by $\omega_{1}$ and $\delta\omega$, while between pulses [Fig.~\ref{Fig1}a(ii)] they undergo $z$ rotations. Over one driving period $T$, the combined protocol yields an effective rotation about $\hat{n}_{\rm eff}$ at frequency $\omega_{\rm eff}$ (Fig.~\ref{Fig1}b)~\cite{maricqThermodynamicsManybodySystems1985b}. Using a Floquet decomposition of the drive dynamics, $U_{\rm drive}(t)=P(t)R_{\hat{n}_{\rm eff}}(\omega_{\rm eff}t)$~\cite{maricqThermodynamicsManybodySystems1985b, ivanovFloquetTheoryMagnetic2021}, the Hamiltonian in the micromotion frame is quantized along $\hat{n}_{\rm eff}$ with an energy gap $\omega_{\rm eff}$,
$\tilde{H}(t)=\omega_{\rm eff}(\hat{n}_{\rm eff}\cdot\vec{I})+\tilde{H}_{\rm dd}(t)$. The static term is removed through a second interaction-frame transformation, after which the residual dipolar terms admit a \emph{bimodal} Fourier expansion,
\begin{equation}
\tilde{H}_{\rm dd}(t)=\sum_{k=-2}^{2}\sum_{n=-\infty}^{\infty} H^{(n,k)}_{\rm dd}\,\Phi_{nk}(t).
\end{equation}
Here the Fourier components $H^{(n,k)}_{\rm dd}$ are weighted by the dynamic phase $\Phi_{nk}(t)=e^{+i (n \omega_d+k \omega_{\rm eff})t}$ (see SI Sec.~\ref{sec:SI_Bimodal}).
\newline\indent
Unlike single-mode Floquet theory, bimodal Floquet theory enables interference between different Fourier modes, giving rise to resonance phenomena. Analogous to the rotating-wave approximation, resonances are characterized by a nearly static phase $\Phi_{n_{0}k_{0}}(t)\simeq 1$, or $n_{0} \omega_d+k_{0} \omega_{\rm eff}\simeq 0$. This leads to a modification of the Floquet-Magnus expansion typically used to describe Floquet heating~\cite{ivanovFloquetTheoryMagnetic2021},
\begin{equation}
\begin{split}
\overline{H}^{F}\simeq H^{(n_{0},k_{0})}_{\rm dd}
-\frac{1}{2}\sum_{n,k} \frac{\left[ H^{(n_{0}-n,k_{0}-k)}_{\rm dd},\,H^{(n,k)}_{\rm dd} \right]}{n \omega_d+k \omega_{\rm eff}}+\cdots ,
\end{split}
\end{equation}
with $(n_{0},k_{0})\neq(n,k)$. At resonance, the Floquet Hamiltonian contains spin-energy-nonconserving terms, leading to rapid heating events enabled by otherwise forbidden multi-photon processes (see SI Sec.~\ref{sec:SI_Res}). With an effective energy gap $\omega_{\rm eff}$, a multi-photon process satisfying $n_{0} \omega_d+k_{0} \omega_{\rm eff}\simeq 0$ implies that $k_{0}$ spins can collectively absorb $n_{0}$ quanta from the drive, conserving the total energy of spins and drive. A triple spin-flip event ($\ket{\uparrow\uparrow\uparrow}\rightarrow\ket{\downarrow\downarrow\downarrow}$) becomes resonant when the driving frequency satisfies $\omega_d\simeq 3\omega_{\rm eff}$ leading to multiple-quantum correlations or operator growth (Fig.~\ref{Fig1}c, orange shaded region). 
\newline\indent
Absorption from the drive is enabled via the dipolar Fourier components $H^{(n_{0},k_{0})}_{\rm dd}$, whose magnitude depends on interspin distances, linking the system’s resilience to multi-photon heating to the positional disorder of the $^{13}$C network. This is illustrated in Fig.~\ref{Fig1}d(ii), which shows the prethermal magnetization extracted after the transient indicated in Fig.~\ref{Fig1}d(i) for $^{13}$C lattice occupancies of 1.1\%, 5\%, and 100\%, spanning disordered to fully ordered configurations. Numerical results represent simulations of a dipolar-coupled 10-spin cluster on a diamond lattice, sampled to match the $^{13}$C occupancy. The quasi-stationary state is calculated by projecting the initial state ($\rho(0)\sim I_{x})$ onto the diagonal ensemble of the one-cycle Floquet propagator~\cite{waughThermodynamicEquilibriumIsolated2004}. To characterize operator growth upon prethermalization, we calculate the average Pauli weight of the prethermal state following Ref.~\cite{schusterOperatorGrowthOpen2023}. Near a multi-photon resonance, heating effects quickly emerge on the transient timescale. This is consistent with the prethermalization hypothesis, $\rho_{\rm pre}\!\sim\!\exp(-\beta \overline{H}^{F})$, where $\beta$ is determined via prethermal energy conservation~\cite{abragam1961principles,maricqThermodynamicsManybodySystems1985b}. Here, $\overline{H}^{F}$ contains additional nonlocal interactions, which generate multiple-quantum correlations. These correlations sharply increase the average Pauli weight of the prethermal state, as seen in Fig.~\ref{Fig1}e around 2.5 and 4.9 kHz, and drive rapid operator growth consistent with the inset cartoons far from and near resonance (Fig.~\ref{Fig1}e(i-ii)). 
\newline\indent
\emph{Resonant heating measurements---}Measurements of prethermal magnetization versus $\delta\omega$, analogous to Fig.~\ref{Fig1}d, are shown in Fig.~\ref{Fig2}a, with the signal averaged over a 10 ms window following the transient period. As expected, only a small dip is observed at the double-spin-flip resonance ($\sim 4.9$ kHz), with negligible changes at the triple-spin-flip resonance ($\sim 2.5$ kHz), indicating disorder-induced protection during the transient period. Notably, this apparent protection at short times does not persist to longer times. We determine the heating rates from the long-time dynamics using a previously established model (see SI Sec.~\ref{sec:SI_fitting} for details). Fig.~\ref{Fig2}b shows the heating rates after the prethermal period, which generally decrease with detuning as the effective energy gap increases. At resonance, however, the heating rates display a sharp peak, indicating additional heating on top of the nonresonant background rate and signaling a breakdown of the anticipated disorder protection. For comparison, Fig.~\ref{Fig2}c shows time traces at $\delta\omega = 0$ and $4.9$ kHz, highlighting accelerated relaxation at resonance. Semi-classical Monte Carlo flip-flop models—including electron-induced relaxation, polarization transport, and resonance offsets—qualitatively reproduce the non-resonant background (see SI Sec.~\ref{sec:SI_MC}) but fail to capture the sharp resonant peaks, which result from many-body effects beyond leading-order flip-flop processes.

\begin{figure}[t]
    \centering
    \includegraphics[width=1\columnwidth]{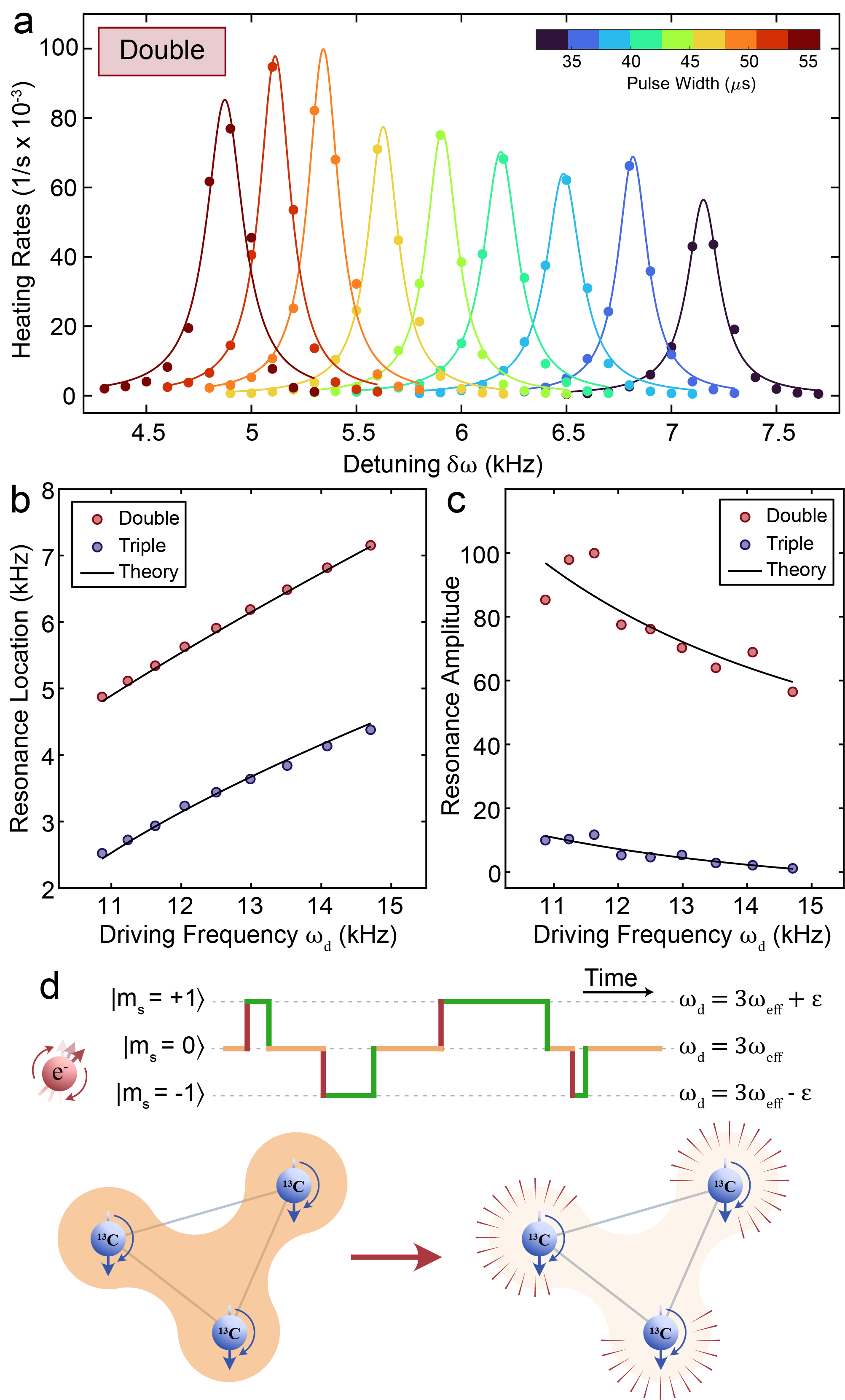}
    \caption{(a) Heating rate vs. detuning near double-spin-flip resonance measured for $\tau_p=32-56$ $\mu$s and fixed inter-pulse spacing of 36 $\mu$s and $\omega_{1}=4.46$ kHz; solid lines are Lorentzian fits. (b) Resonance locations extracted from fits for double- (red) and triple- (blue) spin-flip resonances; black line: theory (see SI Sec.~\ref{sec:SI_Res}). (c) Resonant contribution to heating rate vs. $\omega_d$, compared with expected scaling $\propto\omega_d^{-2}$ (black lines). (d) Schematic of stochastic electron-state switching that modulates resonance condition and dephases higher-order correlations (top: example electron trajectory; bottom: three-spin cluster at resonance).}
    \label{Fig3}
    \zfl{fig3}
\end{figure}

To further investigate these effects, we performed detuning sweeps for various driving frequencies $\omega_d$ by varying the pulse widths of the $\vartheta_x$ pulses while keeping the pulse spacing fixed. Fig.~\ref{Fig3}a tracks the evolution of the double-spin-flip heating rates with detuning; complete datasets are provided in SI Sec.~\ref{sec:SI_ExtData}. Fig.~\ref{Fig3}b shows the resonance locations, whereas Fig.~\ref{Fig3}c shows the resonant heating-rate contributions---given by the portion of the total heating rate above the non-resonant background---extracted from Lorentzian fits. The Lorentzian line shape follows from arguments similar to magnetic $T_{2}$ dephasing in conventional spin systems, with triple- or double-spin-flip resonances playing the role of a Larmor resonance (see SI Sec.\mbox{~\ref{sec:SI_Res}}). The resonance positions are in excellent agreement with theoretical predictions, corresponding to $2\omega_{\rm eff}=\omega_d$ and $3\omega_{\rm eff}=\omega_d$; the full expressions for $\omega_{\rm eff}$ as a function of $\delta\omega$ are provided in SI Sec.~\ref{sec:SI_Drive}.
\newline\indent
\emph{Electron-Mediated Breakdown of Disorder Protection---}The sharp increase in heating rates at resonance is surprising, given the substantial positional disorder in the $^{13}$C network. We attribute this breakdown of disorder-induced protection to the secondary network of electron spins. For simplicity, consider a single electron coupled to a $^{13}$C network. In the absence of electron spin-flip processes, the electronic spin projection $m_{s}$ is conserved since $[H_{\rm hf},S_{z}]=0$. Consequently, the full Floquet dynamics for each $m_{s}$ subspace proceed independently: $\bigoplus_{m_{s}} K_{m_{s}}(t)\exp(-i \overline{H}^{F}_{m_{s}}t)K^{\dagger}_{m_{s}}(0)$, where $K_{m_{s}}(t)$ are kick operators restricted to a particular $m_{s}$ manifold. Fig.~\ref{Fig3}d, however, shows that electron flickering induces stochastic transitions between the different $m_{s}$ manifolds at the electronic relaxation rate $\sim T_{1e}^{-1}$. While Fig.~\ref{Fig3}d illustrates three $m_{s}$ manifolds corresponding to the NV center, the same physical picture applies generically to other paramagnetic defects, including P1 centers. Nuclei experience a stochastic ``switching'' between the corresponding evolution operators (see SI Sec.~\ref{sec:SI_Stochastic}). As shown in the SI, a transition from state $\mu$ to $\nu$ causes a combined kick $K^{\dagger}_{\nu}K_{\mu}$, followed by evolution under $\overline{H}^{F}_{\nu}$ (see SI Sec.~\ref{sec:SI_NonRes}). A kick partially rotates $I_{z}\rightarrow \cos(\epsilon)I_{z}+\sin(\epsilon) I_{\perp}$, with $\epsilon<1$. Between kicks, the perpendicular component is completely dephased by $\overline{H}^{F}_{\nu}$. We distinguish two cases: (i) Away from resonance, $I_{z}$ is largely unaffected because $\overline{H}^{F}_{\nu}$ is mostly energy conserving; heating then proceeds at a rate $T^{-1}_{1}\sim T_{1e}^{-1} \ln[\cos(\epsilon)]$, explaining the decreasing background rate observed in Fig.~\ref{Fig2}b. (ii) At resonance, multi-photon processes are allowed and $\overline{H}^{F}_{\nu}$ is not energy conserving (see SI Sec.~\ref{sec:SI_Res}). Systems with positional disorder are otherwise protected against such mechanisms, as absorption is mediated by the bilinear Fourier components $H^{(n_{0},k_{0})}_{\rm dd}$; the commutator structure of the Floquet-Magnus expansion implies that a collective flip of $k_{0}$ spins can occur no earlier than order $k_{0}-1$. These processes decrease with site occupancy $p$, scaling as $\sim p^{2}d^{2}_{ij}/\omega_d$ for double-spin flips and $\sim p^{3}d_{ij}d_{jk}/\omega_d$ for triple-spin flips, reflecting suppression of higher-order terms by the driving frequency and positional disorder.
\newline\indent
Instead, stochastic hyperfine fields generated by surrounding electrons modulate the resonance condition, enabling decay of the prethermal state even in the presence of positional disorder. This interplay is consistent with the heating rates shown in Fig.~\ref{Fig3}c, which decrease with increasing driving frequency as $\propto\omega_d^{-2}$, in agreement with Fermi’s golden rule, since both processes appear at order $\omega_d^{-1}$ (see SI Sec.~\ref{sec:SI_Res}). Additionally, we repeat measurements similar to Fig.~\ref{Fig2}b under continuous laser illumination, which optically excites the NV center and induces non–spin-conserving transitions through its inter-system crossing, thereby accelerating stochastic switching between different $m_{s}$ manifolds~\cite{selco2025emergent}. Consistent with an electron-mediated mechanism, we observe an enhancement of the resonant heating contribution under illumination (see SI Sec.~\ref{sec:SI_LaserData}).
\newline\indent
\emph{Outlook---}While the role of static on-site disorder in stabilizing exotic dynamical phases such as MBL states has been previously described~\cite{serbynLocalConservationLaws2013a, yaoManyBodyLocalizationDipolar2014a, ponteManyBodyLocalizationPeriodically2015}, our work shows that this stabilization can be significantly disrupted when the on-site disorder fluctuates stochastically. Rather than reinforcing localization, stochastic fluctuations intermittently activate resonant multi-body processes that would otherwise be suppressed by static positional disorder.
\newline\indent
These observations suggest concrete design rules for robust Floquet engineering. Resonance-induced heating can be mitigated by choosing drive parameters that avoid low-order resonance manifolds, by suppressing disorder dynamics (e.g., reducing bath switching rates~\cite{jarmola2012temperature, ethier2017improving} or polarizing/decoupling the environment~\cite{takahashiQuenchingSpinDecoherence2008, belthangady2013dressed,joos2022protecting}), or by engineering pulse trains (e.g. controlled aperiodicity~\cite{moon2025experimental}) that reduce coherent bimodal interference.
\newline\indent
Conversely, the same sharp resonant response can be exploited for DC quantum sensing~\cite{sahin2025micromotion}: tuning close to a resonance makes weak DC fields (or slow parameter drifts) effectively shift the system into resonance, triggering rapid operator spreading and a sudden decay of magnetization. Away from resonance, the system remains in a long-lived prethermal state, while crossing into resonance produces a dramatic breakdown of prethermalization that can serve as a transduction mechanism with gain~\cite{jiang2022floquet}. 
\newline\indent
While our results employ detuned pulsed spin-locking sequences, similar conclusions are expected for more complex periodic drives with multiple intrinsic time scales, which display interference among their frequency components. As a result, we expect analogous resonance-activated breakdowns in other driven quantum systems beyond diamond, including driven dipolar ensembles~\mbox{\cite{martin2023controlling, peng2021floquet}}, solid-state qubits with fluctuating Overhauser or charge environments~\mbox{\cite{bluhm2011dephasing, nichol2015quenching, kuhlmann2013charge}}, and digitally controlled quantum simulators~\mbox{\cite{mullerInteractingTwolevelDefects2015, andersen2025thermalization}}.
\newline\indent
\emph{Acknowledgements.---} We gratefully thank L. J. I. Moon and D. Suter for insightful discussions. We additionally acknowledge funding from ONR (N00014-20-1-2806), AFOSR YIP (FA9550-23-1-0106), and instrumentation support from AFOSR DURIP (FA9550-22-1-0156) and NSF MRI (2320520). CMS acknowledges the NDSEG fellowship.

\bibliographystyle{apsrev4-2}
\bibliography{TSFPaper}

\clearpage
\newpage

\beginsupplement

\begin{center}
{\large\bfseries Supplementary Information for\\
\emph{``Breakdown of Disorder-Suppressed Floquet Heating under Two-Frequency Driving''}}\\[0.5em]
\end{center}

\renewcommand{\tocname}{Contents}
\tableofcontents

\section{Guide to the Supplementary Information}\label{sec:SI_Guide}

This Supplementary Information (SI) provides methodological, theoretical, and extended-data details that support and extend the main text. Throughout, we use the same notation as in the main paper, including the drive frequency $\omega_{d}=2\pi/T$, detuning $\delta\omega$, pulse flip angle $\vartheta=\omega_{1}\tau_{p}$, and the effective single-spin rotation frequency $\omega_{\rm eff}$.

Sec.~\ref{sec:SI_fitting} describes how heating rates are extracted from fits to the decay of the prethermal state. 

Sec.~\ref{sec:SI_Drive} gives the effective drive propagator $U_{\rm drive}(t)$ with explicit expressions for $\{\vartheta_{\rm eff},\varphi_{\rm eff},\omega_{\rm eff}\}$, and states the bimodal resonance condition $n\,\omega_{\rm eff}+k\,\omega_{d}=0$.

Sec.~\ref{sec:SI_MC} presents the semiclassical Monte Carlo model used to simulate electron-induced nuclear relaxation and polarization transport in the dilute $^{13}$C network. We specify the lattice sampling procedure, the electron-centered diffusion barrier, and the relaxation/transport matrices that govern the polarization dynamics. We also explain how resonance-offset effects are incorporated by renormalizing the dipolar and hyperfine couplings using the Floquet micromotion operator. The resulting simulations reproduce key qualitative trends in the detuning dependence (e.g., suppression of dipolar transport near decoupling conditions and the overall decline of relaxation with increasing $\delta\omega$), while clarifying why the sharpest resonance features require many-body physics beyond leading-order Floquet theory.

Sec.~\ref{sec:SI_ToyModel} develops the theoretical framework used to interpret bimodal-Floquet interference and its noise-assisted activation based on a minimal Liouvillian toy model of a few dipolar-coupled nuclei interacting with a relaxing electron and shows that it captures the main resonance phenomenology. Sec.~\ref{sec:SI_Stochastic} casts electron-state switching into a projected (stochastic) Liouville description, making explicit how stochastic transitions generate a sequence of state-dependent Floquet evolutions connected by instantaneous kicks. Sec.~\ref{sec:SI_NonRes} analyzes the non-resonant detuning regime, yielding a kick-induced dephasing picture that sets the smooth background heating/relaxation rate and describing the crossover between dipolar-transport-dominated and hyperfine-shift-dominated limits. Finally, Sec.~\ref{sec:SI_Res} treats the resonant regime, identifying how near-static bimodal Fourier components generate non-energy-conserving multi-spin-flip terms (including double- and triple-spin-flip resonances) and providing estimates for their contributions to the heating rate.

Sec.~\ref{sec:SI_LaserData} and Sec.~\ref{sec:SI_ExtData} provides extended datasets and auxiliary plots underlying the resonance spectroscopy in the main text (including full detuning sweeps at multiple drive frequencies and additional fits), together with any additional numerical results not shown in the main manuscript.

\section{Extraction of Heating Rates}
\label{sec:SI_fitting}
The decay of the prethermal state for $^{13}$C nuclei in diamond coupled to NV and P1 centers, is well described by the product decay law:
\begin{equation}
    \begin{aligned}
        e^{-\sqrt{R_{p}t}}e^{-R_{d}t}.
    \end{aligned}
\end{equation}
The stretched-exponential term with rate $R_{p}$ captures electron-induced relaxation of the nuclear spins, while the monoexponential term with rate $R_{d}$ captures relaxation due to nuclear spin diffusion. This decomposition follows the approach recently introduced in \cite{selco2025emergent} and allows us to separate contributions from electron-mediated and nuclear-spin-mediated processes. Since the resonant heating events we investigate are mediated by electron spins (see main text), we focus on the extracted values of $R_{p}$ as a function of pulse detuning $\delta\omega$. The monoexponential nuclear diffusion term $R_{d}$ is approximately independent of detuning and is not included in the analysis of the resonant heating dynamics.

\section{Effective Drive Dynamics}\label{sec:SI_Drive}

The drive Hamiltonian for off-resonant spin-locking is given by
\begin{equation}
    \begin{aligned}
        H_{\rm drive}(t)=
        \begin{cases}
            \delta\omega I_{z}+\omega_{1} I_{x}&{\rm if}\;0\leq t<\tau_{p}
            \\
            \delta\omega I_{z}&{\rm if}\;\tau_{p}\leq t< T
        \end{cases},
    \end{aligned}
\end{equation}
and extended periodically such that $H_{\rm drive}(t+T)\!=\!H_{\rm drive}(t)$. As a consequence, the drive propagator admits a Floquet form
\begin{equation}
    \begin{aligned}
        U_{\rm drive}(t)
        &=P(t)e^{-i H^{F}_{\rm drive}t}
        \\
        &=P(t)R_{\hat{n}_{\rm eff}}(\omega_{\rm eff}t).
    \end{aligned}
\end{equation}
The second line follows from the fact that the drive propagator represents a pure spin rotation. The micro-motion operator $P(t)$ satisfies
\begin{equation}
    \begin{aligned}
        P(t+T)=P(t), P(0)=\mathbbm{1},
    \end{aligned}
\end{equation}
and may be expressed in terms of a time-dependent Euler rotation
\begin{equation}
    \begin{aligned}
       P(t)&=R_{zyz}[\Lambda(t)]
       \\
       &=R_{z}(\alpha_{t})R_{y}(\beta_{t})R_{z}(\gamma_{t}).
    \end{aligned}
\end{equation}
The Euler angles are periodic $\Lambda_{t+T}=\Lambda_{t}$ in time, and may be determined numerically. For the effective drive Hamiltonian we may choose the axis-angle representation
\begin{equation}
    \begin{aligned}
H^{F}_{\rm drive}=\omega_{\rm eff} (\hat{n}_{\rm eff}\cdot\vec{I}),
    \end{aligned}
\end{equation}
where $\hat{n}_{\rm eff}$  defines a new quantization axis of the system in the micromotion frame
\begin{equation}
    \begin{aligned}
\hat{n}_{\rm eff}=[\sin(\vartheta_{\rm eff})\cos(\varphi_{\rm eff}),\sin(\vartheta_{\rm eff})\cos(\varphi_{\rm eff}),\cos(\vartheta_{\rm eff})].
    \end{aligned}
\end{equation}
The effective axis angles are given by
\begin{equation}
\label{eq:eff_axis}
    \begin{aligned}
&\vartheta_{\rm eff}=\cot^{-1}\{\cos(\gamma/2)\cot(\alpha)+\cot(\psi/2)\csc(\alpha)\sin(\gamma/2)\},
\\
&\varphi_{\rm eff}=-\frac{\gamma}{2},
    \end{aligned}
\end{equation}
with
\begin{equation}
    \begin{aligned}
\psi=\tau_{p}\sqrt{\omega_{1}^{2}+\delta\omega^{2}},
\;\gamma=(T-\tau_{p})\delta\omega,
\;\alpha=\tan^{-1}(\omega_{1},\delta\omega).
    \end{aligned}
\end{equation}
Similarly, the effective frequency $\omega_{\rm eff}$ is given by
\begin{equation}
    \begin{aligned}
\omega_{\rm eff}=\cos^{-1}[
&c(\gamma)c(\psi)
+c^{2}(\alpha)(1+c(\gamma)c(\psi))
\\
+&s^{2}(\alpha)(c(\gamma)+c(\psi))
-2c(\alpha)c(\gamma)c(\psi)-1]/T,
    \end{aligned}
\end{equation}
where $c(x)=\cos(x)$ and $s(x)=\sin(x)$. Treating $\omega_{\rm eff}$ as a function of $\delta\omega$, the resonance conditions in the offset parameter are found numerically from
\begin{equation}
    \begin{aligned}
k_{0}\omega_{d}-n_{0}\omega_{\rm eff}(\delta\omega)=0.
    \end{aligned}
\end{equation}

\section{Semiclassical Monte Carlo Model}\label{sec:SI_MC}

\begin{figure}[b]
    \centering
    \includegraphics[width=\columnwidth]{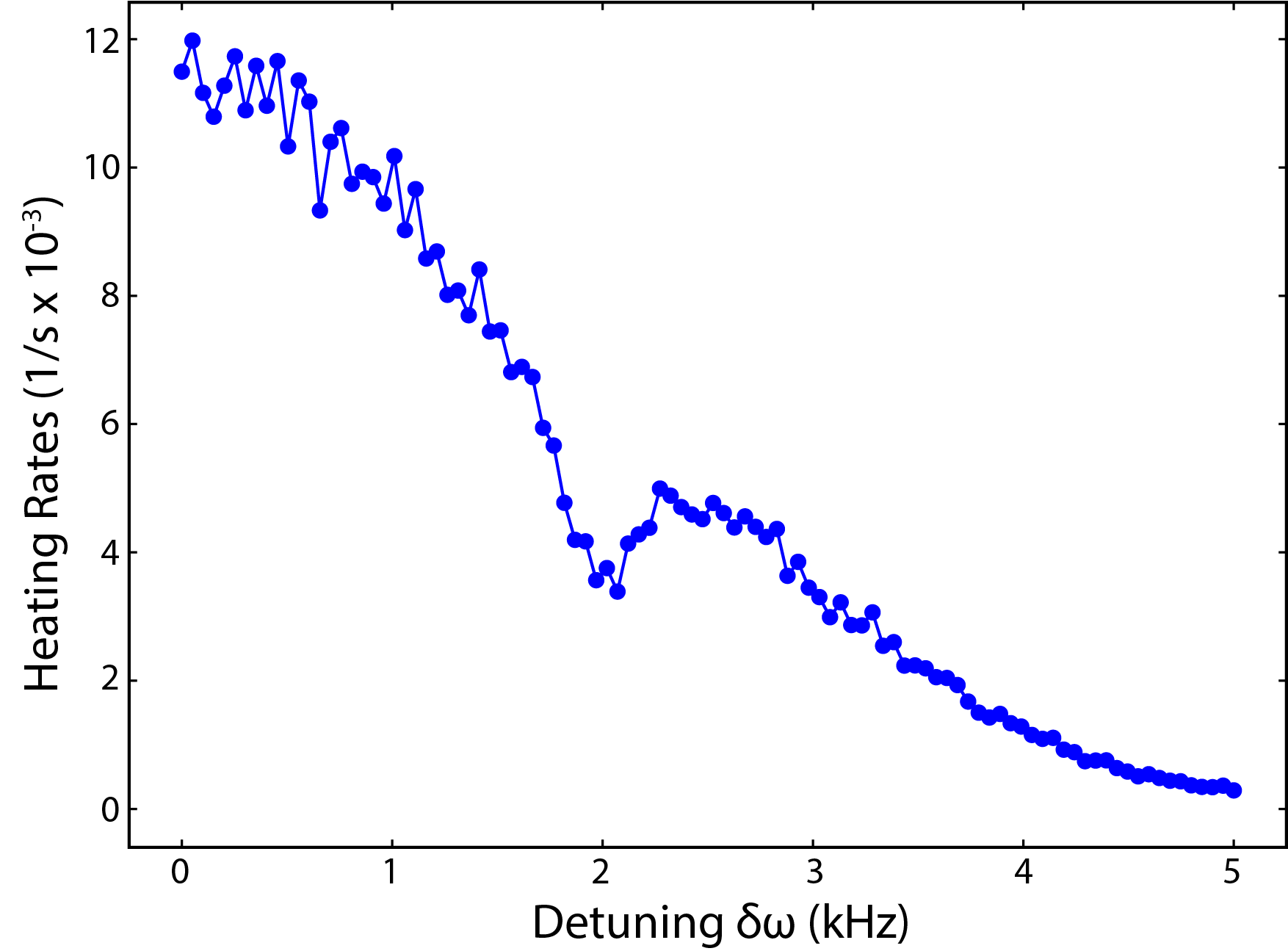}
    \caption{Semiclassical Monte Carlo simulations of nuclear polarization decay rates as a function of detuning $\delta\omega$. The model reproduces the overall decrease of heating with increasing detuning. A weak reduction near $\delta\omega\approx2.3$ kHz is attributed to suppressed spin diffusion when strongly relaxing nuclei near paramagnetic impurities become dynamically decoupled from their neighbors. Sharp resonance peaks observed experimentally are not captured, as they arise from many-body double- and triple-spin-flip processes beyond first-order Floquet theory and beyond the scope of the semiclassical model.}
    \label{fig:OffsetMonteCarlo}
\end{figure}

Qualitative features of the off-resonant heating behavior can be captured within a semi-classical Monte Carlo framework. Polarization transport in the $^{13}$C spin network is modeled as a Markovian hopping process arising from dipolar-mediated pairwise spin flip–flop events, in which two spins exchange polarization while conserving total magnetization. Electron-induced relaxation is incorporated as an on-site stochastic relaxation process. Lattice sites consistent with the diamond crystal structure are generated and populated with $^{13}$C nuclei and electron spins using binomial statistics at the specified concentrations. A 16 Å spin-diffusion barrier is imposed around each electron, excluding $^{13}$C spins within this radius from further calculations. The polarization dynamics for a given lattice configuration follow
\begin{equation}
\dot{p}(t) = (R + W)p(t),
\end{equation}
where $p(t)$ is a vector containing the polarization of each nuclear spin. The matrices $R$ and $W$ are responsible for electron-induced relaxation and polarization transport, respectively. The relaxation matrix $R$ is given by
\begin{equation}
R_{ij} = -\eta \sum_{\mu=1}^{N_e} h_{i\mu}^2 \delta_{ij},
\end{equation}
where $\delta_{ij}$ is the Kronecker delta, $\eta$ is an effective model parameter to account for correlation time of the electrons and the finite number of electrons in the simulation volume, whereas $h_{i\mu}$ is the hyperfine coupling constant between nuclear spin $i$ and paramagnetic impurity $\mu$, given by
\begin{equation}
h_{i\mu}=-\frac{\mu_{0}}{4\pi}\frac{\hbar\gamma_{\rm e}\gamma_{\rm n}}{r^{3}_{i\mu}}\left(3\cos^{2}\!\vartheta_{i\mu} - 1\right),
\end{equation}
where $\mu_{0}$ is the permeability of free space, $\hbar$ is the reduced Planck's constant, $\gamma_{\rm e}$ and $\gamma_{\rm n}$ are the gyromagnetic ratios of the electron and $^{13}$C respectively, $r_{i\mu}$ is the distance between nuclear spin $i$ and paramagnetic impurity $\mu$, and $\vartheta_{i\mu}$ is the angle between the inter-spin vector and the external magnetic field. The off-diagonal elements of the dipolar transport matrix $W$ are given by
\begin{equation}
W_{ij} = d_{ij}^2\kappa^{2}J(0),
\end{equation}
whereas diagonal elements are given by
\begin{equation}
W_{ii} = -\sum_{i\neq j} W_{ij},
\end{equation}
ensuring conservation of the total polarization. The dipolar coupling constant $d_{ij}$ between nuclear spins $i$ and $j$ is given by
\begin{equation}
d_{ij}=-\frac{\mu_{0}}{4\pi}\frac{\hbar\gamma^{2}_{\rm n}}{r^{3}_{ij}}\left(3\cos^{2}\!\vartheta_{ij} - 1\right).
\end{equation}
The spectral density $J(\omega)$ quantifies the energy overlap of two spins~\cite{abragam1961principles}. For simplicity, we assume that all nuclei have the same resonance frequency and treat $J(0)$ as a model parameter. 

To account for resonance offset effects, we employ rescaled dipolar couplings $d_{ij}$ and hyperfine couplings $h_{i\mu}$ derived from the standard high-frequency expansion of the Floquet Hamiltonian. For this purpose, we work in the micro-motion frame defined by the micro-motion operator $P(t)$ and retain only secular parts with respect to the quantization axis. The resulting scaled coupling constants can be expressed as follows
\begin{equation}
\label{eq:coupling_scaling}
\begin{aligned}
d_{ij}&\leftarrow d_{ij}\sum_{m=-2}^{+2}T^{-1}\int_{0}^{T}d^{2}_{0m}(\vartheta_{\rm eff})e^{+i m \varphi_{\rm eff}}D^{2}_{m0}{[}\Lambda(t){]}\,dt,
\\
h_{i\mu}&\leftarrow h_{i\mu}\sum_{m=-1}^{+1}T^{-1}\int_{0}^{T}d^{1}_{0m}(\vartheta_{\rm eff})e^{+i m \varphi_{\rm eff}}D^{1}_{m0}{[}\Lambda(t){]}\,dt.
\end{aligned}
\end{equation}
Here, $\vartheta_{\rm eff}$ and $\varphi_{\rm eff}$ parameterize the effective quantization axis, $d^{l}_{mn}$ are reduced Wigner matrix elements, and $D^{l}_{m0}{[}\Lambda(t){]}$ are full Wigner matrix elements parametrized by the micro-motion Euler angles $\Lambda(t)$. 

A proxy for the experimentally observable nuclear polarization is obtained by taking a configurational average of $p(t)$:
\begin{equation}
\overline{p}(t) = \langle p(t)\rangle_{\rm conf}.
\end{equation}
As discussed in detail in our previous work~\cite{selco2025emergent}, the configurationally averaged polarization $\overline{p}(t)$ follows the universal decay form
\begin{equation}
\overline{p}(t) \sim \exp(-R_{d}t)\exp(-R_{p}t),
\end{equation}
applicable even under detuned Floquet driving. Here, $R_{p}$ reflects a \emph{direct} (``paramagnetic'') heating contribution from coupling to the fluctuating electron bath, whereas $R_{d}$ captures an \emph{indirect} heating contribution mediated by polarization transport through the nuclear network toward electron-centered relaxation sinkholes. Figure~\ref{fig:OffsetMonteCarlo} shows numerically extracted values of $R_{p}$ for pulse-sequence parameters matched to the experiment. Notably, the model accurately reproduces the monotonic decline of the non-resonant background heating with increasing $\delta\omega$, in agreement with the expression for the rescaled hyperfine couplings. The small decrease in $R_{p}$ near $\delta\omega=2.3$ kHz can be understood in terms of the dipolar “magic angle” condition: when a strongly relaxing spin near a paramagnetic impurity becomes effectively decoupled from its neighbor, it can no longer drain polarization from surrounding spins. As a result, nearby nuclei retain polarization for longer times, reducing the net heating rate. In practice, this effect is expected to be washed out due to slight resonance frequency differences. Most notably, the semiclassical simulations do not capture the sharp resonance peaks observed experimentally. This is expected, as these peaks arise from intrinsically many-body effects that go beyond first-order Floquet theory. Semiclassical spin-diffusion models, in contrast, are inherently based on two-spin flip–flop processes and thus only describe the dynamics of local observables, such as the polarization of individual spins.

\section{Theory}\label{sec:SI_theory}

\subsection{Minimal toy model}\label{sec:SI_ToyModel}

A minimal model for electron-induced Floquet heating under different detuning conditions is given by a system of $N_{I}$ dipolar coupled nuclei, interacting with a single electron. The electron is assumed to be coupled to the lattice; to fully account for this coupling, we adopt a Liouville-space formalism
\begin{equation}
\frac{d}{dt}\rho(t)=\mathcal{L}\rho(t).
\end{equation}
The Liouvillian, $\mathcal{L}=\mathcal{L}_{0}+\mathcal{D}$, contains unitary ($\mathcal{L}_{0}$) and dissipative contributions ($\mathcal{D}$). The unitary dynamics of the toy system are divided into internal interactions
\begin{equation}
\mathcal{L}_{\rm int}=\mathcal{L}^{\rm NN}_{\rm dd}+\mathcal{L}^{\rm NE}_{\rm hf},
\end{equation}
and the drive part acting on the nuclei
\begin{equation}
\mathcal{L}_{\rm drive}(t)=\omega_{1}(t)\mathcal{L}^{\rm N}_{x}+\delta\omega\,\mathcal{L}^{\rm N}_{z}.
\end{equation}
The driving frequency is given by $\omega_{d}=2\pi/T$. The nuclei interact via their secular dipolar interaction
\begin{equation}
\begin{aligned}
\mathcal{L}_{\rm dd}^{\rm NN}&=-i \sum_{i<j}^{N_{I}}d_{ij}\hat{T}^{ij}_{20},
\\
T^{ij}_{20}&=3 I_{i}^{z}I_{j}^{z}-{\bf I}_{i}\cdot{\bf I}_{j}.
\end{aligned}
\end{equation}
The hat indicates promotion to a commutation superoperator
\begin{equation}
\hat{T}^{ij}_{20}Q={[}T^{ij}_{20},Q{]}.
\end{equation}
The indices indicate a spherical tensor of rank $(l,m)=(2,0)$ with respect to spatial rotations of the nuclei, $R(\Omega)T_{lm}R^{\dagger}(\Omega)=T_{ln}D^{l}_{nm}(\Omega)$. The hyperfine interaction is given by
\begin{equation}
\mathcal{L}_{\rm hf}^{\rm NE}=-i \sum_{i=1}^{N_{I}}h_{i}\hat{T}^{i}_{10}\hat{S}_{z}.
\end{equation}
The hyperfine interaction is a spherical tensor of rank $(l,m)=(1,0)$ with respect to rotation of the nuclei. We split the hyperfine interaction as follows
\begin{equation}
\mathcal{L}_{\rm hf}^{\rm NE}
=\sum_{m_{s}}\mathcal{L}^{\rm N}_{m_{s}}\hat{P}^{\rm E}_{m_{s}}=-i \sum_{i=1}^{N_{I}}\sum_{m_{s}}h^{m_{s}}_{i}\hat{T}^{i}_{10}\hat{P}^{\rm E}_{m_{s}},
\end{equation}
emphasizing the electron state-dependent hyperfine shift ($h^{m_{s}}_{i}=m_{s}h_{i}$). The dissipative coupling of the electron to the lattice is captured by a Lindblad dissipator, assumed to evoke $T_{1}$ and $T_{2}$ relaxation processes
\begin{equation}
\mathcal{D}^{\rm E}Q=\gamma\sum_{j=1}^{3}L_{j}Q L^{\dagger}_{j}-\frac{1}{2}\{ L^{\dagger}_{j}L_{j},Q\},
\end{equation}
with
\begin{equation}
L_{1}=S_{z}, \quad L_{2}=\frac{1}{\sqrt{2}}S_{+}, \quad L_{3}=\frac{1}{\sqrt{2}}S_{-}.
\end{equation}
The total evolution of the toy model is governed by
\begin{equation}
\mathcal{L}(t)=\mathcal{L}_{\rm int}+\mathcal{L}_{\rm drive}(t)+\mathcal{D}^{\rm E}.
\end{equation}

\begin{figure}[t]
    \centering
    \includegraphics[trim={0 13cm 0 0},clip,width=\columnwidth]{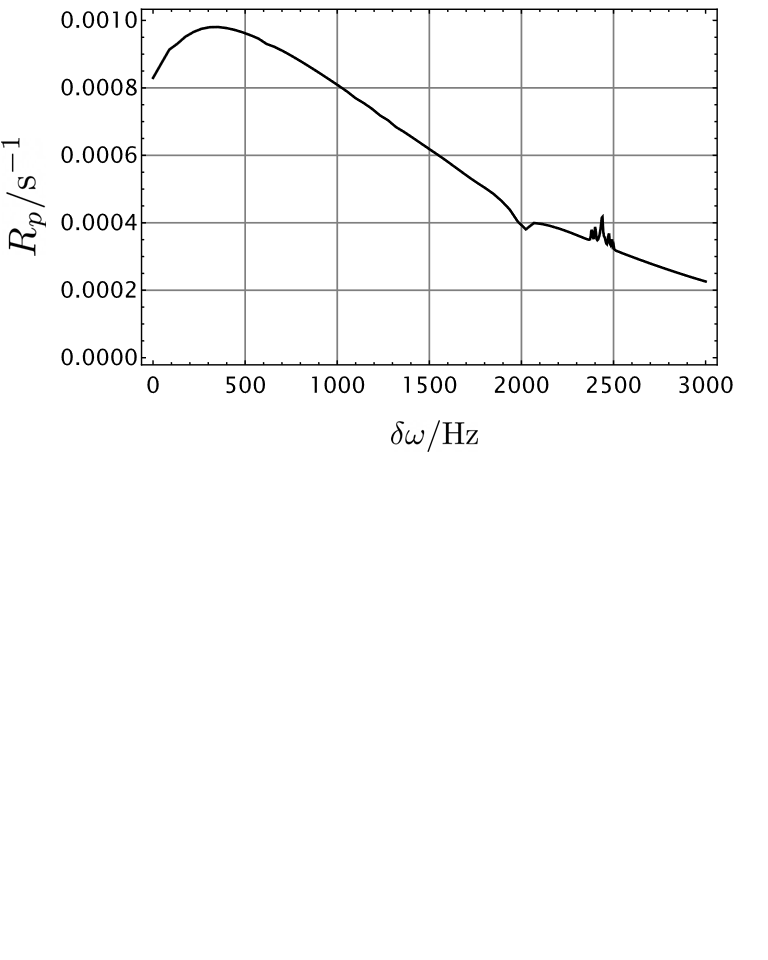}
    \caption{Numerically simulated heating rates are shown for a minimal model comprising three nuclei and one electron under detuned pulsed spin-locking. The results represent an average over 20 randomly generated three-spin clusters confined within a 10 nm sphere sampled from the diamond lattice. For each configuration, the heating rates are further averaged over electron positions distributed on a 10 point Lebedev grid located 40 nm from the center of the three-spin cluster. The pulse sequence parameters are chosen in agreement with experimental details: $\vartheta_{x}=\pi/2$, $\tau_{p}=56\;\mu{\rm s}$, and $T=92\;\mu{\rm s}$, $T_{1e}=50$ ms.}
    \label{fig:rate_average}
\end{figure}

As a concrete example, Fig.~\ref{fig:rate_average} shows simulations of the Floquet heating rate under detuned pulsed-spin locking for a minimal model of three nuclei and a NV center, focusing on the triple-spin flip condition. The numerically simulated rates reproduce the qualitative trends observed in the experimental data shown in Fig.~\ref{Fig2}b. Notably, the heating rates exhibit a pronounced increase near $\delta\omega\simeq2.5$ kHz, which is in agreement with triple-spin-flip processes. The slight decrease in the paramagnetic rate around 2 kHz may be attributed to a first-order decoupling condition of the nuclear dipolar interactions. This feature matches the semi-classical simulations shown in Fig.~\ref{fig:OffsetMonteCarlo}. As explained above, at the magic angle condition individual spins relax independently and the decay of the total polarization is dominated by the longest relaxing nucleus. In contrast to the semi-classical model, however, the toy model also captures the non-classical double- and triple-spin flip processes.

\subsection{Bimodal Floquet theory}\label{sec:SI_Bimodal}

The decoherence dynamics of the toy model may be analyzed in more detail within the framework of bimodal Floquet theory. In a frame co-rotating with the drive, interactions experience modulations at both frequencies $\omega_{d}$ and $\omega_{\rm eff}$, resulting in a bimodal Floquet problem
\begin{equation}
\begin{aligned}
\tilde{\mathcal{L}}(t)
&=S^{\dagger}_{\rm drive}(t)\mathcal{L}(t)S_{\rm drive}(t)-\mathcal{L}_{\rm drive}(t)
\\
&=\sum_{n=-2}^{+2} \sum_{k=-\infty}^{+\infty}\mathcal{L}^{(n,k)}_{\rm int}\,\Phi_{nk}(t)+\mathcal{D}^{\rm E},
\end{aligned}
\end{equation}
with the Fourier phase $\Phi_{nk}(t)$ given by
\begin{equation}
\Phi_{nk}(t)=e^{+i n \,\omega_{d}t}e^{+i k \,\omega_{\rm eff}t}.
\end{equation}
The bimodal decomposition of the Liouvillian may be evaluated by first moving into the micro-motion frame
\begin{equation}
\begin{aligned}
P^{\dagger}(t)\mathcal{L}(t)P(t)&-P^{\dagger}(t)\dot{P}(t)
\\
&=\omega_{\rm eff}(\hat{n}_{\rm eff}\cdot\vec{\mathcal{L}})+P^{\dagger}(t)\mathcal{L}_{\rm int}P(t)+\mathcal{D}^{\rm E}.
\end{aligned}
\end{equation}
We may now tilt the frame so that: $\hat{n}_{\rm eff}\cdot\vec{\mathcal{L}}\rightarrow \mathcal{L}_{z}$. The dipolar terms, for example, are then given by
\begin{equation}
\begin{aligned}
\hat{T}^{ij}_{20}\rightarrow\hat{T}^{ij}_{2k}D^{2}_{km}(\vartheta_{\rm eff},\varphi_{\rm eff})D^{2}_{m0}(\Lambda_{t}).
\end{aligned}
\end{equation}
A subsequent transformation into a frame defined by $\omega_{\rm eff}\mathcal{L}_{z}$ introduces a dynamical phase factor
\begin{equation}
\begin{aligned}
\hat{T}^{ij}_{20}\rightarrow\hat{T}^{ij}_{2k}D^{2}_{km}(\vartheta_{\rm eff},\varphi_{\rm eff})D^{2}_{m0}(\Lambda_{t})e^{+ik\omega_{\rm eff}t}.
\end{aligned}
\end{equation}
The Wigner matrix elements may now be expanded as a Fourier series in $\omega_{d}$ due to the periodicity of the Euler angles 
\begin{equation}
\begin{aligned}
\hat{T}^{ij}_{20}\rightarrow\hat{T}^{ij}_{2k}D^{2}_{km}(\vartheta_{\rm eff},\varphi_{\rm eff})f^{l}_{mn}e^{+i(k\omega_{\rm eff}+n\omega_{d})t},
\end{aligned}
\end{equation}
with the Fourier coefficients 
\begin{equation}
f^{l}_{mn}=T^{-1}\int_{0}^{T}D^{l}_{m0}{[}\Lambda(t){]}e^{-i n \omega_{d}t}\,dt.
\end{equation}
Similar results hold for the hyperfine interaction. Resonant processes occur whenever 
\begin{equation}
n_{0} \omega_{d}+k_{0} \omega_{\rm eff}\simeq 0,
\end{equation}
leading to a static dynamical phase. In the absence of any dissipative dynamics, an effective dynamical description, taking into account possible resonance effects, may be constructed via a bimodal high-frequency expansion~\cite{scholzOperatorbasedFloquetTheory2010a,ivanovFloquetTheoryMagnetic2021}
\begin{equation}
\tilde{S}_{\rm int}(t)\simeq \exp(K(t))\exp(\mathcal{L}_{\rm eff}t)\exp(K^{\dagger}(0)),
\end{equation}
with the effective Liouvillian $\mathcal{L}_{\rm eff}$ and the kick operator $K(t)$ given by
\begin{align}
\mathcal{L}_{\rm eff}&\simeq \mathcal{L}^{(1)}_{\rm int}+\mathcal{L}^{(2)}_{\rm int},
\\
\mathcal{L}^{(1)}_{\rm int}&=\sum_{(n_{0},k_{0})}\mathcal{L}_{\rm int}^{(n_{0},k_{0})},
\\
\mathcal{L}^{(2)}_{\rm int}&=-\frac{1}{2}\sum_{(n_{0},k_{0})}\sum_{(n_{0},k_{0})\neq(n,k)}\frac{{[}\mathcal{L}_{\rm int}^{(n_{0}-n,k_{0}-k)},\mathcal{L}_{\rm int}^{(n,k)}{]}}{n\omega_{\rm eff}+k\omega_{\rm d}},
\\
K(t)&\simeq \sum_{(n,k)\neq (n_{0},k_{0})}\frac{\mathcal{L}_{\rm int}^{(n,k)}}{n\omega_{\rm eff}+k\omega_{\rm d}}\,\Phi_{nk}(t).
\end{align}

\subsection{Stochastic Liouville formulation}\label{sec:SI_Stochastic}

The population subspace $\mathbf{P}=\{P^{\rm E}_{+1},P^{\rm E}_{0},P^{\rm E}_{-1}\}$ of the electron is invariant under the action of $\mathcal{L}_{0}(t)$, since the driving field only acts on the nuclei. Projecting the Liouvillian onto $\mathbf{P}$ we obtain
\begin{equation}
\begin{aligned}
\mathcal{L}^{\mathbf{P}}(t)&=\sum_{m_{s},m^{'}_{s}}\{w_{m_{s}m^{'}_{s}}+\delta_{m_{s}m^{'}_{s}}\mathcal{L}^{\mathbf{P}}_{m_{s}}(t)\}\vert P^{\rm E}_{m_{s}})(P^{\rm E}_{m^{'}_{s}} \vert,
\end{aligned}
\end{equation}
where we have introduced the state-dependent generators $\mathcal{L}^{\mathbf{P}}_{m_{s}}(t)$
\begin{equation}
   \mathcal{L}^{\mathbf{P}}_{m_{s}}(t)=\mathcal{L}^{\rm N}_{\rm drive}(t)+\mathcal{L}^{\rm NN}_{\rm dd}+\mathcal{L}^{\rm N}_{m_{s}}
\end{equation}
The projected generator is formally equivalent to a stochastic Liouville equation, with the nuclei experiencing a varying hyperfine potential $\mathcal{L}^{\rm N}_{m_{s}}$ due to electron flickering. For a single NV center, for example, $w$ is given by
\begin{equation}
w_{m_{s}m^{'}_{s}}=T^{-1}_{1e}
\left[\begin{array}{ccc}
     -1 & 1 & 0
     \\
     1 & -2 & 1
     \\
     0 & 1 & -1
\end{array}\right].
\end{equation}
The reduced state of the nuclei $\rho_{\rm N}(t)$ can thus be thought of as a noise functional over the electron flickering
\begin{equation}
\rho_{\rm N}(t)=\int_{X(t)}\rho_{\rm N}(t\vert X(t))\mathcal{D}{[}X(t){]},
\end{equation}
where $X(t)$ is a given sample path.

For P1 and NV centers, the electron flickering time \mbox{$T_{1e}\sim {\rm ms}$} is significantly longer than the driving period (\mbox{$T_{1e}/T\sim 10$}). On average, the electron will thus sojourn in its state for several driving cycles before flickering out. As a first approximation, we perform a high-frequency expansion of the state-dependent generators' contributions, and subsequently reintroduce $w_{m_{s}m^{'}_{s}}$. After alignment of the co-rotating coordinate system with the effective quantization axis, the Fourier series of the state-dependent generator is given by
\begin{equation}
\label{eq:L_state_dep_fourier}
\begin{aligned}
\tilde{\mathcal{L}}^{\mathbf{P}}_{m_{s}}(t)
=&\sum_{n,k}(\mathcal{L}_{2}^{(n,k)}+\mathcal{L}_{1}^{(n,k)})\Phi_{nk}(t)
\\
=&\sum_{n,k}\left(\sum_{i<j}d_{ij}\sum_{m} \hat{T}^{ij}_{2n}D^{2}_{nm}(\vartheta_{\rm eff},\varphi_{\rm eff})f_{mk}^{2}\right)\Phi_{nk}(t)
\\
&+\sum_{n,k}\left(\sum_{i} h^{m_{s}}_{i}\sum_{m}\hat{T}^{i}_{1n}D^{1}_{nm}(\vartheta_{\rm eff},\varphi_{\rm eff})f_{mk}^{1}\right)\Phi_{nk}(t).
\end{aligned}
\end{equation}

\subsection{Non-resonant detuning effects}\label{sec:SI_NonRes}

\begin{figure}[t]
    \centering
    \includegraphics[width=1\columnwidth]{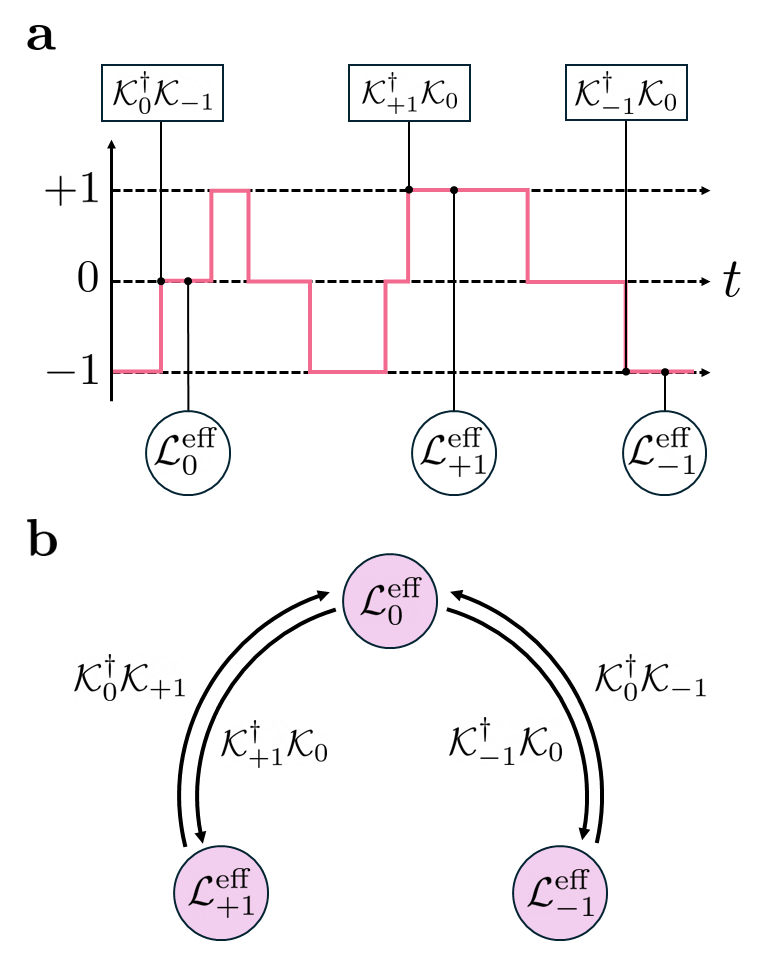}
    \caption{Stochastic interpretation of the electron-driven Floquet heating for the case of an NV center. {\bf a)} Sample trajectory (pink) as the electron undergoes transitions between its Zeeman states. While the electron sojourns in a state $m_{s}$, the spins evolve under $\mathcal{L}^{\rm eff}_{m_{s}}$. Upon transition to a state $\nu$ the spins experience the instantaneous ``kick'' $\mathcal{K}^{\dagger}_{m'_{s}}\mathcal{K}_{m_{s}}$, before continuing their evolution under $\mathcal{L}^{\rm eff}_{m^{'}_{s}}$. {\bf b)} When averaged over many realizations, the dynamics of the spins may be described within the stochastic Liouville formalism in which the nuclei undergo a compound Markov process.}
    \label{fig:markov_jumps}
\end{figure}

In the non-resonant case (\mbox{$ n_{0}\omega_{\rm eff}+k_{0}\omega_{d}\neq 0$} for any Fourier pair $(n_{0},k_{0})\neq (0,0)$), the (electron-state dependent) kick operator is approximately given by
\begin{equation}
\label{eq:K_state_dep}
\tilde{\mathcal{K}}_{m_{s}}(t)
\simeq\sum_{n,k\neq0}\sum_{l,l'=1}^{2}\frac{\mathcal{L}_{l}^{(n,k)}}{n\omega_{\rm eff}+k\omega_{\rm d}}\Phi_{nk}(t),
\end{equation}
whereas the effective Liouvillian, to second order, is given by
\begin{equation}
\label{eq:L_state_dep_eff}
\tilde{\mathcal{L}}^{\rm eff}_{m_{s}}
\simeq\sum_{l=1}^{2}\mathcal{L}_{l}^{(0,0)}-\frac{1}{2}\sum_{n,k\neq0}\sum_{l,l'=1}^{2}\frac{{[}\mathcal{L}_{l}^{(-n,-k)},\mathcal{L}_{l'}^{(n,k)}{]}}{n\omega_{\rm eff}+k\omega_{\rm d}}.
\end{equation}
The leading-order term, $\mathcal{L}_{2}^{(0,0)}$, captures the dominant dipolar interactions driving polarization exchange. In contrast, $\mathcal{L}_{1}^{(0,0)}$ accounts primarily for the residual hyperfine shifts. The second-order terms represent corrections to the polarization dynamics and hyperfine shifts.

\begin{figure}[t]
    \centering
    \includegraphics[trim={0 0 0 0},clip,width=\columnwidth]{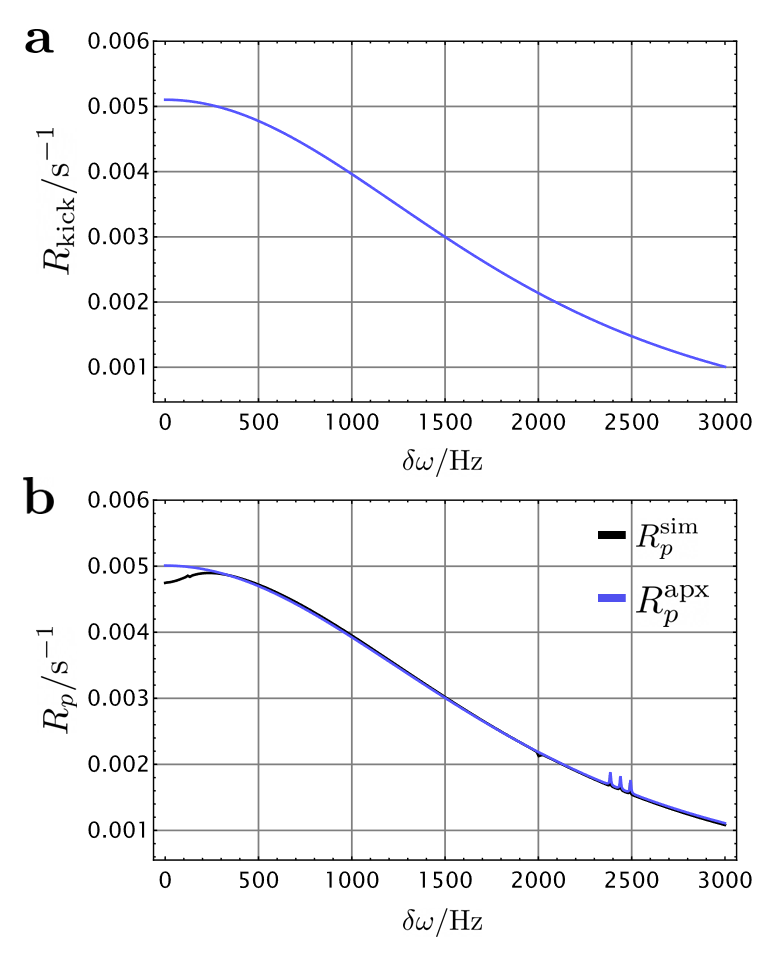}
    \caption{a) Kick dephasing contribution (Eq.~\ref{eq:R_approx_kick}) for a random three-spin cluster for pulsed spin-locking as a function of the detuning. b) Total heating rates as a function of detuning, accounting for both kick dephasing and resonant multi-photon contributions. Black line indicates numerically simulated results, whereas the blue line represents the approximation given by Eq.~\ref{eq:R_approx_tot}. The pulse sequence parameters are: $\vartheta_{x}=\pi/2$, $\tau_{p}=56\;\mu{\rm s}$, and $T=92\;\mu{\rm s}$, $T_{1e}=50$ ms. The nuclei are placed at: $r_{1}=a_{0}(1,1,-3)/4$, $r_{2}=a_{0}(4,2,2)/4$, $r_{3}=a_{0}(-2,-4,2)/4$, whereas the electron is situated at $r_{e}=a_{0}(0.2,-0.2,20.25)$. Here, $a_{0}=357$ pm denotes the diamond lattice constant.}
    \label{fig:rate_kickdephase}
\end{figure}

It is straightforward to verify that, for the non-resonant case, $\tilde{\mathcal{L}}^{\rm eff}_{m_{s}}$ commutes with $\mathcal{L}^{\rm N}_{z}$. This would imply the conservation of $I_{z}$ even in the presence of electron flickering. A more complete picture of the heating dynamics is displayed in Fig.~\ref{fig:markov_jumps}a and highlights the role of the state-dependent kick operators $\tilde{\mathcal{K}}_{m_{s}}$. Within the stochastic Liouville framework, while the electron sojourns in state $m_{s}$, the system continues to evolve under the effective Liouvillian $\tilde{\mathcal{L}}^{\rm eff}_{m_{s}}$. Upon transition from state $m_{s}$ to $m^{'}_{s}$, the system experiences an instantaneous frame realignment, $\tilde{\mathcal{K}}^{\dagger}_{m^{'}_{s}}\tilde{\mathcal{K}}_{m_{s}}$, after which it continues to evolve under $\tilde{\mathcal{L}}^{\rm eff}_{m^{'}_{s}}$. The frame transition $\tilde{\mathcal{K}}^{\dagger}_{m^{'}_{s}}\tilde{\mathcal{K}}_{m_{s}}$ generates, in general, a small orthogonal polarization
\begin{equation}
e^{\tilde{\mathcal{K}}^{\dagger}_{m'_{s}}}e^{\tilde{\mathcal{K}}_{m_{s}}}I_{z}\simeq \cos(\epsilon)I_{z}+\sin(\epsilon)I_{\perp},
\end{equation}
with $\epsilon\ll1$ representing a small rotation. A first approximation to the heating process may then be given as follows. Prior to each transition, we assume that the orthogonal polarization is fully dephased due to differential hyperfine shifts and dipolar interactions. In contrast, the $z$-component remains preserved due to the commutativity of the effective generators. After $n$ transitions the averaged polarization is then approximately given by
\begin{equation}
\cos(\epsilon)^{n}\sim \exp\!\left(T^{-1}_{1e} \,\ln[\cos(\epsilon)]t\right),
\end{equation}
and the electron-induced decay rate for the kick-dephasing approximation is given by
\begin{equation}
\label{eq:R_approx_kick}
R_{\rm kick}\simeq -T^{-1}_{1e}\ln[\cos(\epsilon)].
\end{equation}
Fig.~\ref{fig:rate_kickdephase} shows the behavior of $R_{\rm kick}$ as a function of the detuning $\delta\omega$. A comparison with Fig.~\ref{fig:OffsetMonteCarlo} shows that $R_{\rm kick}$ displays the expected qualitative features of the electron contribution:
\\
\noindent{\bf Regime 1)} For small detuning ($\delta\omega \approx 0$), the effective axis is closely aligned with the $x$ axis ($\vartheta_{\rm eff}\simeq \pi/2$). Eqs.~\ref{eq:L_state_dep_fourier} and~\ref{eq:L_state_dep_eff} then indicate that the first-order hyperfine contribution vanishes, $\mathcal{L}^{(0,0)}_{1}=0$, and the coherent dynamics are mainly dominated by dipolar terms. The absence of $\mathcal{L}^{(0,0)}_{1}$ implies that the hyperfine contributions are concentrated into the non-static Fourier components with $n\neq 0$, leading to strong kick dynamics. In between kicks, dipolar interactions and second-order hyperfine shifts dephase orthogonal components resulting in a pronounced polarization decay.

\noindent{\bf Regime 2)} For nonzero detuning frequencies, the polar angle $\vartheta_{\rm eff}$ of the effective quantization axis tilts out of the $x$--$y$ plane toward the $z$ axis. At a specific intermediate detuning, a condition analogous to the magic-angle case is satisfied, such that $\vartheta_{\rm eff} = \cos^{-1}(1/\sqrt{3})+\delta$. Eqs.~\ref{eq:L_state_dep_fourier} and~\ref{eq:L_state_dep_eff} then indicate that the first-order dipolar contribution vanishes, $\mathcal{L}^{(0,0)}_{2}=0$. In this scenario, each nucleus relaxes independently at a rate proportional to its individual distance from the electron, causing the total polarization to persist for at least as long as the longest nuclear relaxation time. This leads to an apparent reduction in the overall polarization decay rate. Such behavior is not fully captured by the kick-dephasing approximation, which implicitly assumes local equalization via dipolar polarization transport.
\\
\noindent {\bf Regime 3)} For large detuning $\delta\omega\gg 1$, the effective axis asymptotically aligns with the $z$ axis, such that $\vartheta_{\rm eff} \simeq 0$. Eqs.~\ref{eq:L_state_dep_fourier} and~\ref{eq:L_state_dep_eff} then indicate that most of the hyperfine interaction enters the first-order contribution $\mathcal{L}^{(0,0)}_{1}\neq 0$. Consequently, only small orthogonal hyperfine interactions contribute to the non-static Fourier components with $n\neq 0$, and kick dynamics are suppressed. This allows the total polarization to persist over significantly extended timescales.

\subsection{Resonant detuning effects}\label{sec:SI_Res}

In the resonant case, non-zero Fourier pairs $\{(n_{0},k_{0})\}$ can display static Fourier phases ($\Phi_{n_{0}k_{0}}(t)\simeq 0$). The first-order and second-order effective generators then pick up non-energy-conserving contributions
\begin{equation}
\begin{aligned}
\tilde{\mathcal{L}}^{\rm eff}_{m_{s}}
\simeq&\sum_{(n_{0},k_{0})}\sum_{l=1}^{2}\mathcal{L}_{l}^{(n_{0},k_{0})}
\\
&-\frac{1}{2}\sum_{(n_{0},k_{0})}\sum_{(n_{0},k_{0})\neq(n,k)}\sum_{l,l'=1}^{2}\frac{{[}\mathcal{L}_{l}^{(n_{0}-n,k_{0}-k)},\mathcal{L}_{l'}^{(n,k)}{]}}{n\omega_{\rm eff}+k\omega_{\rm d}}.
\end{aligned}
\end{equation}
For the case of triple spin-flip resonances \mbox{$\omega_{d}=3 \omega_{\rm eff}$}, for example, the resonant Fourier pairs satisfy
\begin{equation}
n_{0}=-3 k_{0}.
\end{equation}
Since the first Fourier index ranges from $-2$ to $+2$, a triple spin-flip resonance cannot affect the first-order contribution, but it introduces triple spin-flip (3SF) processes at second order
\begin{equation}
\begin{aligned}
\tilde{\mathcal{L}}^{\rm 3SF}_{m_{s}}=-\frac{1}{2}\sum_{k,l,l^{'}}&\frac{{[}\mathcal{L}_{l}^{(-2,1-k)},\mathcal{L}_{l'}^{(-1,k)}{]}}{(3k-1)\omega_{\rm eff}}
+\frac{{[}\mathcal{L}_{l}^{(-1,1-k)},\mathcal{L}_{l'}^{(-2,k)}{]}}{(3k-2)\omega_{\rm eff}}
\\
+&\frac{{[}\mathcal{L}_{l}^{(+2,1-k)},\mathcal{L}_{l'}^{(+1,k)}{]}}{(3k+1)\omega_{\rm eff}}
+\frac{{[}\mathcal{L}_{l}^{(+1,1-k)},\mathcal{L}_{l'}^{(+2,k)}{]}}{(3k+2)\omega_{\rm eff}}.
\end{aligned}
\end{equation}
In general, for spin-1/2 systems, $n$ spin-flip processes can occur no earlier than order $n-1$ in the high-frequency expansion. This observation reflects the bilinear structure of the dipolar Hamiltonian and the fact that each commutation operation can increase the number of interacting spins by at most one (${[}{[}ij,jk{]},jl{]}\rightarrow {[}ijk,kl{]}\rightarrow ijkl$).

The 3SF processes manifest as a sharp increase in the heating rate centered around the nominal resonance condition $\delta\omega_{\rm 3SF}$, implicitly defined by
\begin{equation}
\frac{\omega_{\rm eff}(\delta\omega_{\rm 3SF})}{T}-\frac{2\pi}{3}=0.
\end{equation}
This nominal resonance condition, however, experiences a slight modification due to a state-dependent hyperfine shift
\begin{equation}
\delta\omega^{m_{s}}_{\rm 3SF}\simeq \delta\omega_{\rm 3SF}-\langle h^{m_{s}}\rangle.
\end{equation}
Here, $\langle h^{m_{s}}\rangle=1/N_{I}\sum_{i}h^{m_{s}}_{i}$ represents the mean hyperfine shift while the electron sojourns in state $m_{s}$. 

Consider, for example, the coupling to a single NV center, which exhibits three distinct resonance conditions. If the detuning value matches one of these conditions, for example $\delta\omega=\delta\omega^{+1}_{\rm 3SF}$, the nuclei participate in triple spin-flip events. While the electron remains in the state $m_{s}=+1$, triple spin flips generate non-stationary multiple-quantum correlations that evolve under the effective Zeeman interaction. The evolution is stochastically interrupted by transition events, which, in addition to kick-induced dephasing, further enhance the triple spin-flip decoherence process. As an example, the heating rate contributions from triple-spin-flip events at the central resonance condition $\delta\omega\sim\delta\omega^{0}_{\rm 3SF}$, can be understood by transforming to a resonant frame~\cite{maricqThermodynamicsManybodySystems1985b}, defined by
\begin{equation}
\mathcal{L}^{0}_{\rm res}=\omega_{1}(t)\mathcal{L}_{x}+\delta\omega^{0}_{\rm 3SF} \mathcal{L}_{z},
\end{equation}
by introducing a fictitious detuning matched to $\delta\omega^{0}_{\rm 3SF}$. The resonant propagator is cyclic with cycle time \mbox{$T_{c}=3T$}, $S^{0}_{\rm res}(T_{c})=\mathbbm{1}$. This makes the problem amenable to single-mode Floquet theory, with the addition of a small residual Zeeman interaction ($\vert\delta\omega -\delta\omega^{0}_{\rm 3SF}\vert \lesssim 1$). A straightforward, if somewhat tedious, matrix perturbation analysis shows that the triple-spin-flip contribution to the heating rate roughly follows
\begin{equation}
R^{0}_{\rm 3SF}\sim \frac{T^{-1}_{1e}}{(3  T^{-1}_{1e})^{2}+(2 \delta\omega^{0}_{\rm 3SF})^{2}},
\end{equation}
which represents a Lorentzian centered at $\delta\omega^{0}_{\rm 3SF}$. If the resonance conditions are sufficiently far apart, the triple spin-flip contribution may be approximated by a superposition
\begin{equation}
\label{eq:R_approx_tot}
R_{\rm 3SF}\simeq \sum_{m_{s}} R_{\rm 3SF}^{m_{s}}.
\end{equation}
A comparison of $R^{\rm apx}_{p}$ against numerical results for one particular three-spin cluster is shown in Fig.~\ref{fig:rate_kickdephase}b. Apart from small deviations at $\delta\omega\simeq 0$, the approximate heating rates agree well with numerically simulated results. Noticeably, $R^{\rm apx}_{p}$ captures the sharp increase in the decay rate around the triple spin-flip ($\delta\omega\simeq 2.5$) kHz. A similar analysis can be carried out for the other resonance conditions. Since these conditions are generally well separated, an approximate expression for the total heating rate for our toy model, accounting for both double- and triple-spin-flip processes, is given by
\begin{equation}
\label{eq:R_approx_tot_v2}
R^{\rm apx}_{p}\simeq R_{\rm kick}+\sum_{m_{s}} R_{\rm 2SF}^{m_{s}} +\sum_{m_{s}} R_{\rm 3SF}^{m_{s}}.
\end{equation}

\section{Accelerated Resonant Heating Under Laser Illumination}\label{sec:SI_LaserData}

\begin{figure}[t]
    \centering
    \includegraphics[width=\columnwidth]{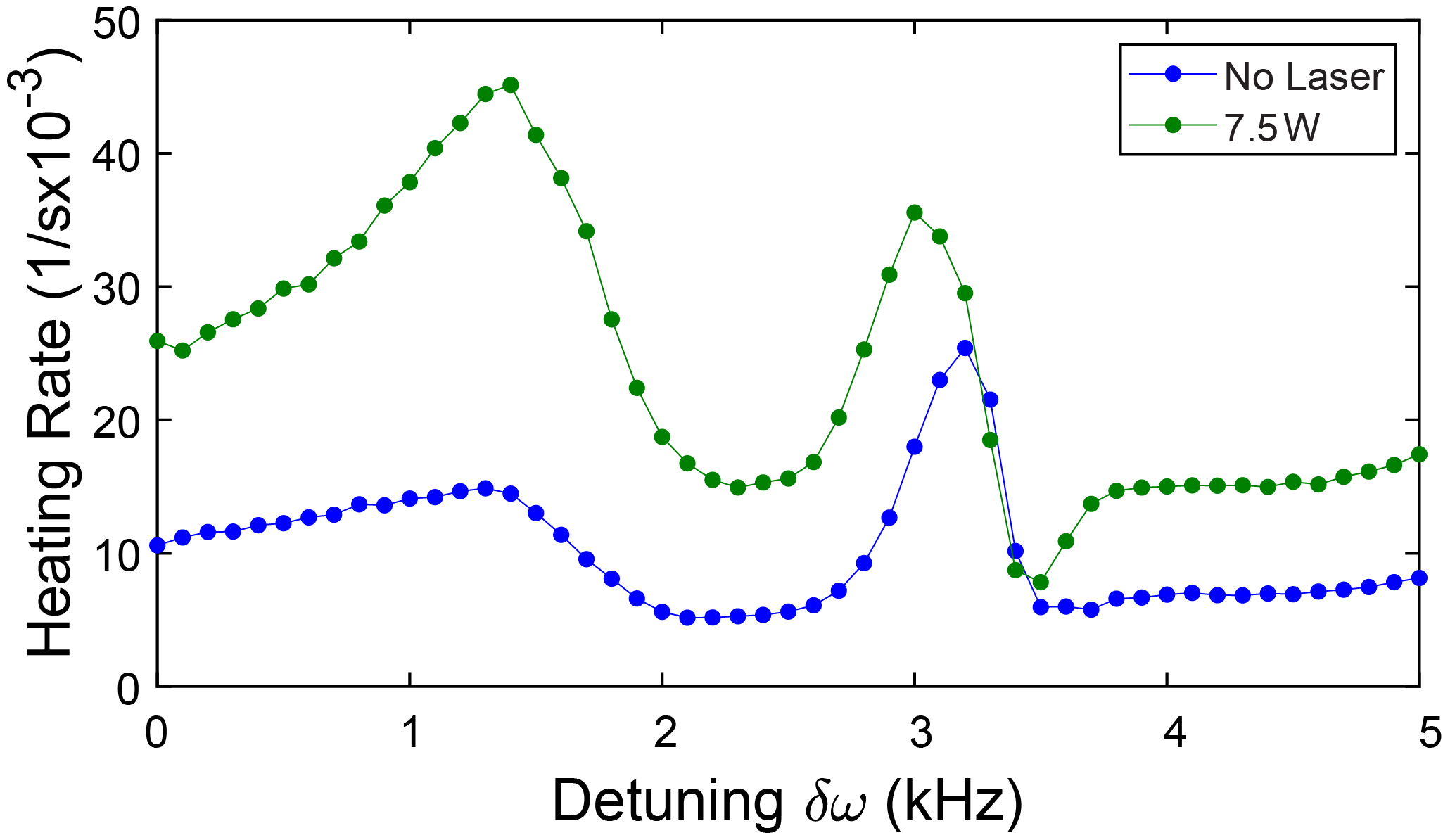}
    \caption{Effect of laser illumination on Floquet heating near the triple–spin–flip resonance. Long-time heating rate as a function of detuning $\delta\omega$ measured at 100 K with pulse width 40 $\mu$s and pulse spacing 40 $\mu$s, shown for measurements performed without laser illumination (blue) and with 7.5 W of 532 nm laser power applied continuously during the Floquet sequence (green). The triple–spin–flip resonance occurs near $\delta\omega\approx3.1$ kHz, where laser illumination leads to a pronounced enhancement of the heating rate, consistent with an electron-mediated activation of multi-spin-flip processes.}
    \label{fig:TSF_LaserData}
\end{figure}

To further investigate the microscopic origin of the multi-spin-flip resonances observed in the main text, we examine the effect of optical illumination during the Floquet sequence. This experiment is motivated by the fact that the observed double- and triple-spin-flip resonances are activated by electron-spin dynamics. Under continuous optical illumination, the NV center is repeatedly excited to its electronic excited state and undergoes non–spin-conserving transitions through its inter-system crossing. These processes effectively shorten the electron correlation time~\cite{selco2025emergent}, resulting in faster stochastic transitions between the different electronic spin projections $m_{s}$. Since both the resonant and off-resonant Floquet heating mechanisms rely on stochastic switching between the different $m_{s}$ manifolds (see SI Sec.~\ref{sec:SI_NonRes},~\ref{sec:SI_Res}), accelerating these fluctuations is expected to enhance both the resonant and off-resonant contributions to the long-time heating rate.

These measurements were performed at a reduced temperature of 100 K, where the electronic spin polarization is enhanced and the electron spin populations are more sensitive to optical pumping. At this temperature, laser illumination is expected to more strongly perturb the electronic steady state, while leaving the intrinsic nuclear dipolar couplings unchanged. The Floquet driving parameters were fixed to a pulse width of 40 $\mu$s and a pulse spacing of 40 $\mu$s, corresponding to a triple–spin–flip resonance near$\delta\omega\approx3.1$ kHz.
We first measured the detuning dependence of the long-time heating rate in the absence of laser illumination. We then repeated the entire experiment while illuminating the sample with 7.5 W of 532 nm laser power continuously during the Floquet evolution. As shown in Fig.~\ref{fig:TSF_LaserData}, laser illumination leads to a pronounced enhancement of the heating rate at the triple–spin–flip resonance, as well as away from the resonance condition. This observation provides strong evidence that the triple-spin-flip process is electron-mediated, as optical excitation primarily affects the NV center electronic spin populations and their stochastic switching dynamics, rather than directly modifying nuclear–nuclear interactions.

Within the framework developed in SI Sec.~\ref{sec:SI_theory}, this behavior is naturally explained by laser-induced modifications to the electron spin dynamics, where laser illumination causes faster switching statistics between electronic spin manifolds though inter-system crossing. These effects increase the probability that small nuclear clusters are intermittently tuned in and out of the bimodal Floquet resonance condition, thereby enhancing noise-assisted multi-spin-flip absorption.

In addition to the resonant enhancement near $\delta\omega\approx3.1$ kHz, Fig.~\ref{fig:TSF_LaserData} also shows a gradual increase in the heating rate at low detunings ($\delta\omega\lesssim1.5$ kHz) for both laser-on and laser-off conditions. Notably, this feature is absent in the room-temperature datasets presented in the main text, suggesting that it may be specific to the low-temperature regime explored here. A detailed investigation of this low-detuning enhancement is beyond the scope of the present work and will be the subject of future studies.


\section{Extended Data}\label{sec:SI_ExtData}

\subsection{Triple Spin Flip Resonances}

\begin{figure}[t]
    \centering
    \includegraphics[width=\columnwidth]{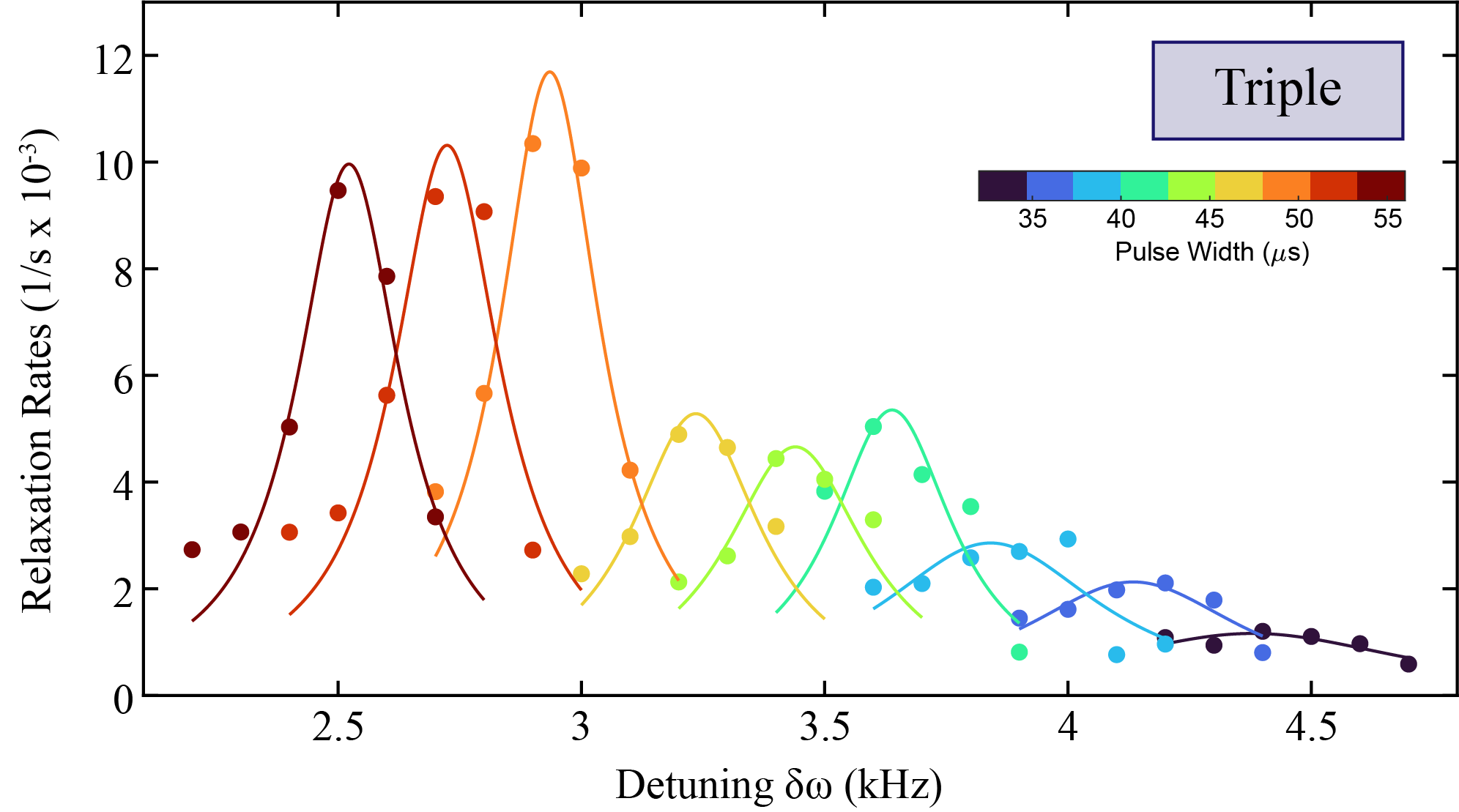}
    \caption{Heating rate versus detuning $\delta\omega$ measured near the triple–spin–flip resonance for multiple driving frequencies $\omega_d$, obtained by varying the pulse length (33–56 $\mu$s) at fixed interpulse delay (36 $\mu$s). Solid lines are Lorentzian fits used to extract the resonance positions and contributions.}
    \label{fig:TSF_Resonances}
\end{figure}

In this section we present in Fig.~\ref{fig:TSF_Resonances} detuning sweeps of the heating rate near the triple–spin–flip resonance for multiple driving frequencies $\omega_d$, analogous to the double–spin–flip data shown in Fig. 3a of the main text. The measurements are performed using the same pulse sequence and analysis protocol, with $\omega_d$ varied by adjusting the pulse length while keeping the inter-pulse delay fixed. Compared to the double–spin–flip case, the triple–spin–flip resonances are weaker, consistent with theory. Nevertheless, resonance peaks are observed in the long-time heating rates when the bimodal resonance condition $3\omega_{\rm eff}=\omega_d$ is satisfied. The resonance locations and contributions are again extracted from Lorentzian fits to the data.

\begin{figure*}[t]
    \centering
    \includegraphics[width=\textwidth]{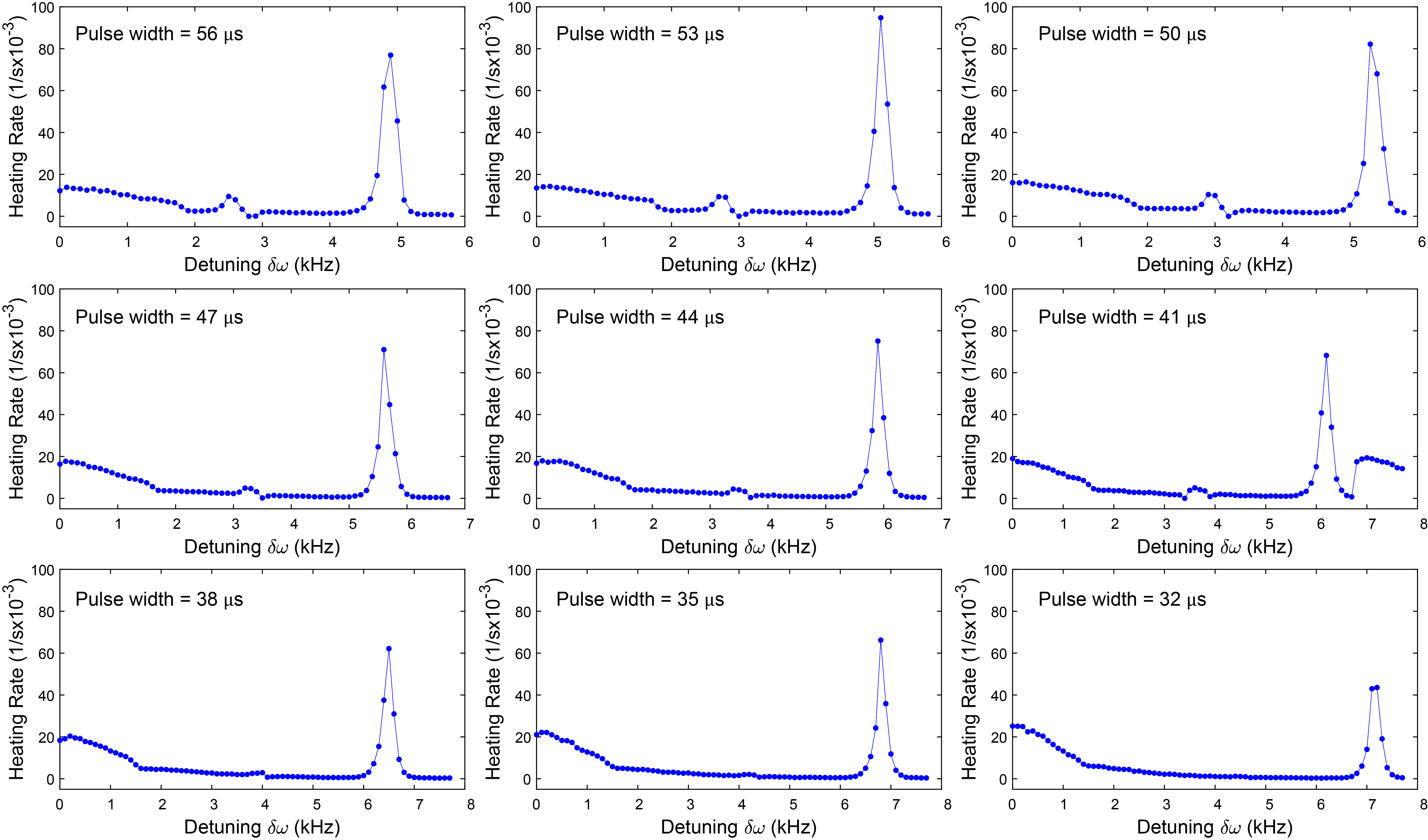}
    \caption{Full detuning dependence of the fitted stretched-exponential rate $R_{p}$ (blue) due to electron-induced relaxation and exponential rate $R_{d}$ (red) due to nuclear spin diffusion for nine driving frequencies $\omega_{d}$ corresponding to different pulse widths with the inter-pulse delay fixed at 36 $\mu$s. Pronounced blue peaks correspond to the data shown in Fig. 3a in the main text.}
    \label{fig:Extended_Data}
\end{figure*}

\subsection{Full Detuning Dependence for All Driving Frequencies}

For each detuning sweep and driving frequency $\omega_d$, the long-time decay of the prethermal magnetization is fit to the form: $M(t)=e^{-\sqrt{R_{p}t}}e^{-R_{d}t}$, where $R_d$ captures relaxation driven by nuclear spin diffusion within the $^{13}$C network, while $R_p$ reflects electron-induced relaxation arising from stochastic hyperfine field fluctuations. This product form, consisting of exponential and stretched-exponential contributions, was developed and validated in Ref.~\cite{selco2025emergent}. In the main text, we focus on the detuning dependence of $R_p$, as the resonance-enhanced heating associated with double- and triple–spin–flip processes manifests predominantly in this stretched-exponential contribution. 
Here we present the full detuning sweeps for each of the nine driving frequencies shown in Fig. 3. 

\end{document}